\documentclass[11pt,a4paper,preprintnumbers,nofootinbib]{revtex4}

\pdfoutput=1

\usepackage[T1]{fontenc}
\usepackage[utf8]{inputenc}
\usepackage{amsmath}
\usepackage{amssymb}
\usepackage{hyperref}
\usepackage[final]{graphicx} 

\usepackage{color}

\begin{document}

\title{Clues for flavor from rare lepton and quark decays}
\author{Ivo \surname{de Medeiros Varzielas}}
\email{ivo.de@soton.ac.uk}
\affiliation{School of Physics and Astronomy, University of Southampton, Southampton, SO17 1BJ, U.K.}
\author{Gudrun Hiller}
\email{ghiller@physik.uni-dortmund.de}
\affiliation{Fakultät für Physik, TU Dortmund, Otto-Hahn-Str.4, D-44221 Dortmund, Germany}
\begin{abstract}
Flavor symmetries successfully explain lepton and quark masses and mixings yet it is usually hard to distinguish different models that predict the same mixing angles.
Further experimental input could be available, if the agents of flavor breaking are sufficiently low in mass and detectable or
if new physics with non-trivial flavor charges is sufficiently low in mass and detectable. The recent hint for lepton-nonuniversality in  the ratio of branching fractions
$B \to K \mu \mu$ over $B \to K e e$, $R_K$,
suggests the latter, at least for indirect detection via rare decays. We demonstrate the discriminating power of the rare decay data on flavor model building taking into account
viable leptonic mixings and show
how correlations with other observables exist in leptoquark models. We give expectations for branching ratios $B \to K \ell \ell^\prime, B_{(s)} \to \ell \ell^\prime$ and
$\ell \to \ell^\prime \gamma$, and Higgs decays $h \to \ell \ell^\prime$. 
\end{abstract}

\preprint{DO-TH 15/02, QFET-2015-04}

\maketitle

\section{Introduction}

The LHCb collaboration \cite{Aaij:2014ora} has presented recent data on lepton-nonuniversality (LNU): $R_K$, the ratio of
branching fractions $B \to K \mu \mu$ over $B \to K e e$, $R_K$ has been measured as
\begin{align} \label{eq:RKdata}
R_{K}^{\rm LHCb} =0.745 \pm^{0.090}_{0.074} \pm 0.036
\end{align}
in the dilepton invariant mass squared bin $1 \, \mbox{GeV}^2 \leq q^2 < 6  \, \mbox{GeV}^2$, which represents a deviation of $2.6 \sigma$ from lepton universality \cite{Hiller:2003js}.
In view of its generic leptoquark interpretation \cite{Hiller:2014yaa} 
we investigate directions in flavor model building that link the electron-muon non-universality to other flavor observables particularly lepton masses and mixing and lepton-flavor violation (LFV).
In general if LNU is present, LFV can be expected, too \cite{Glashow:2014iga}. 
$R_K$ has also been addressed  within (composite) leptoquark models \cite{Gripaios:2014tna}; 
they do not aim at explaining LFV.
Also recently, \cite{Sahoo:2015wya} considered scalar leptoquark contributions to rare leptonic $B$ meson decays.

Here we consider frameworks where both phenomena (LNU, LFV) stem from the same (scalar) sector.
To be more specific, we investigate the possibility of simultaneously accommodating LNU as observed through $R_K$, 
and LFV, as observed in neutrino mixing and as hinted at 2.5 $\sigma$ by CMS in $h \to \tau \mu$ \cite{Khachatryan:2015kon}.
We do so by adapting frameworks with non-Abelian flavor symmetries that predict the leptonic mixing matrix.

Due to the Standard Model (SM) gauge group, the maximal flavor symmetry is an $U(3)$ for each chiral quark and lepton species.
This observation motivates the use of $SU(3)_F$ or subgroups. After discussing briefly a framework of leptoquarks within the continuous group $SU(3)_F$, \cite{deMedeirosVarzielas:2005ax}
we consider explicit realizations with the discrete group $A_4$ in frameworks within supersymmetry (SUSY) \cite{Altarelli:2005yx, Altarelli:2009kr, Varzielas:2010mp,Varzielas:2012ai}. While the specific results are related to the details of the models, our general conclusion 
is generic even if other discrete subgroups of $SU(3)_F$ are invoked.
If the same flavor symmetry breaking fields, often referred to as flavons,
are involved in Higgs and leptoquark couplings, the respective Yukawa structures are related and there are links between what is predicted for LNU and LFV (in lepton mixing and Higgs decays). 

The paper is organized as follows: 
in Section \ref{sec:patterns} we identify viable, data-driven patterns for flavor structure of the leptoquark to fermions coupling matrix.
Predictions for rare LFV decays of leptons and $B$-mesons are given in Section \ref{sec:predictions}.  The results of these sections are rather independent of flavor models.
In Section \ref{sec:FS} these benchmarks are compared to what is realized in theoretical flavor models,
where we consider models based on $SU(3)_F$ and $A_4$ known for explaining lepton flavor. Rare Higgs
decays are addressed in Section \ref{sec:higgs}.  In Section \ref{sec:conclusion} we conclude. In several appendices we give formulae and auxiliary information.  

\section{Data-driven flavor patterns \label{sec:patterns}}

We consider models with a scalar $SU(2)$ doublet leptoquark $\Delta(3,2)_{1/6}$, with electric charges $(2/3,-1/3)$  and one with a scalar $SU(2)$ triplet leptoquark $\Delta(3,3)_{-1/3}$, with electric charges $(2/3, -1/3, -4/3)$. Such frameworks correspond to the RL model and the LL model, respectively, of \cite{Hiller:2014yaa}, which discussed them as ``single particle'' explanations  of $R_K$. For this reason, apart from having renamed the leptoquarks as $\Delta$, throughout this section we use the notation from \cite{Hiller:2014yaa}.

In view of the  uncertainties in the current data in the following we 
neglect factor of $1/2$ differences in the semileptonic 4-fermion operators after Fierz-ordering and other small differences between the RL and LL model variants.
This way the phenomenological constraints on the leptoquark parameters relevant to our study  give identical hierarchies in flavor patterns, which we hence study jointly.
We stress that despite the present day patterns being similar, there are differences in terms of model-building. The $SU(2)$-doublet leptoquark model couples to right-handed quarks
\footnote{The RL model has similarities with the R-parity violating minimal supersymmetric standard model (MSSM); the corresponding term is the $\lambda^\prime L Q d^c$ coupling.
Note also that the $\Delta(3,2)_{1/6}$ with mass much lighter than the GUT-scale does not require further model-building
to avoid proton decay \cite{Dorsner:2012nq}.
}
\begin{align}
\label{eq:L_dLQ}
\mathcal{L}=-\lambda_{q \ell}\, \Delta\, (\bar{q} P_L \ell)  +h.c.\ ,
\end{align}
while the  $SU(2)$-triplet leptoquark model couples to left-handed quarks
\begin{align}
\label{eq:L_tLQ}
\mathcal{L}=-\lambda_{q \ell}\, \Delta^*\, (\bar{q}^{*} \ell) +h.c.\ .
\end{align}
In each case,  the couplings to the leptons are left-chiral and quark $(q)$ and lepton $(\ell)$ denote flavor (generation) indices. We denote the mass of the leptoquark by $M$.

We continue discussing constraints on the generalized Yukawa 
$\lambda_{q\ell}$ for $q=d,s,b$ and $\ell=e,\mu,\tau$: 
\begin{eqnarray}
 \lambda \equiv
\left( \begin{array}{ccc}
\lambda_{d e} &  \lambda_{d \mu}  & \lambda_{d \tau}\\
\lambda_{s e} &  \lambda_{s \mu}  & \lambda_{s \tau}\\
\lambda_{b e}  &  \lambda_{b \mu} & \lambda_{b \tau} 
\end{array} \right)  \, .
\end{eqnarray}
$R_K$ implies at 1 $\sigma$  \cite{Hiller:2014yaa}:
\begin{align} \label{eq:RKbound}
0.7 \lesssim & {\rm  Re}[  \lambda_{se}\lambda_{be}^*  - \lambda_{s \mu}\lambda_{b\mu}^* ]  \frac{(24 {\rm TeV})^2 }{M^2} \lesssim 1.5 \, .
\end{align}
We restrict our analysis to $M \gtrsim 1$ TeV to be conservative with collider bounds \cite{Agashe:2014kda,CMS:2014qpa,ATLAS:2013oea,WICHMANN:2013voa}, hence, in view of Eq.~(\ref{eq:RKbound}), $|\lambda_{se}\lambda_{be}^*-\lambda_{s \mu}\lambda_{b\mu}^*| \gtrsim 2 \cdot 10^{-3}$.

There is also a contribution to $B_s$ mixing from a box diagram with leptons and leptoquarks in the loop giving rise to an operator of the form $\bar{b}s\bar{b}s$ with the complex coefficient $ (\sum_\ell \lambda_{s\ell}\lambda_{b \ell}^*)^2/(16 \pi^2 M^2)$.
Numerically, 
\begin{align} \label{eq:Bsmixbound}
(\lambda_{se}\lambda_{be}^*+\lambda_{s\mu}\lambda_{b\mu}^*+\lambda_{s\tau}\lambda_{b \tau}^*)^2 \lesssim \frac{M^2 }{(12 \, \mbox{TeV} )^2} \, , \quad \left(  \frac{M^2 }{(17.3 \, \mbox{TeV} )^2} \right) \, ,
\end{align}
where the weaker bound is obtained from the mass difference $|\Delta m_s^{\rm NP}/\Delta m_s^{\rm SM}| \lesssim 0.15 $, whereas the stronger one in parentheses stems from the upper bound on the $B_s -\bar B_s$ mixing phase $-0.015 \pm 0.035$ (defined relative to the SM phase in the $b \to c \bar c s$ decay amplitude) \cite{Amhis:2014hma}.
The complementarity of the  $|\Delta B|=1$ and $|\Delta B|=2$ constraints allows to fix the mass scale and the dimensionless couplings separately,
yielding  an upper limit $M \lesssim {\cal{O}}(50)$ TeV.
Couplings to different quark flavors are within four orders of magnitude, 
$10^4 \gtrsim   \lambda_{s\ell}/\lambda_{b\ell} \gtrsim  10^{-4}$  \cite{Hiller:2014yaa}.

The simplest scenarios are when the leptoquark couples to one generation of leptons only,
as considered in \cite{Hiller:2014yaa}. In this case  $ \lambda$ reads, 
 \begin{eqnarray} \label{eq:single}
\lambda^{[e]} \equiv \left( \begin{array}{ccc}
 \lambda_{d e} &  0  & 0\\
\lambda_{s e} &  0  & 0\\
\lambda_{b e}  &  0 & 0
\end{array}
\right)  \, , \quad
\lambda^{[\mu]} \equiv \left( 
\begin{array}{ccc}
0 &  \lambda_{d \mu}  & 0\\
0 &  \lambda_{s \mu}  & 0\\
0  &  \lambda_{b \mu} & 0
\end{array}
 \right)  \, .
\end{eqnarray}
A third variant with couplings to $\tau$ only cannot explain non-universality between electrons and muons as hinted currently by $R_K$ data.
With either of those limits, $\lambda^{[e]}$ or $\lambda^{[\mu]}$, there is no LFV in rare processes.
\footnote{Generically $R_K<1$ can be explained by a suppressed branching ratio to muons, or an enhanced one to electrons, or combinations thereof, the details
of which driven by an underlying model of flavor.
Present data, however, is much more copious for rare decays into muons than into electrons due to the large data samples from the hadron machines at Fermilab and CERN LHC. This is likely to change in the nearer term future with the Belle II experiment taking data. 
As a consequence at present the constraints on the $b \to s ee$ modes are much more loose. In addition, hints for new physics in $B \to K^* \mu \mu$ angular distributions could point to a simultaneous explanation with $R_K$, see \cite{Blake:2015tda} and references therein. So, the rather minimal scenario where $R_K$ originates at least predominantly from muon mode suppression and the electrons are SM-like presently has phenomenological support.}

Generalizing the single lepton scenario we consider the  hierarchical pattern
\begin{eqnarray}  \label{eq:hierarchy}
 \lambda^{[\rho \kappa]} \sim \lambda_0\left( 
\begin{array}{ccc}
\rho_d \kappa &  \rho_d   & \rho_d  \\
\rho \kappa &  \rho   & \rho  \\
\kappa &  1 & 1
\end{array} 
\right)  \, .
\end{eqnarray}
Here, we allow for quark flavor suppressions $\rho_d =\lambda_{d \ell}/\lambda_{b \ell}$ and $\rho =\lambda_{s \ell}/\lambda_{b \ell}$ 
having larger couplings for heavier generations. This is suggested by the observed quark mass pattern; see Appendix \ref{app:FN} for details on how this  is  realized with the Froggatt-Nielsen (FN) mechanism \cite{Froggatt:1978nt} from $U(1)$-flavor symmetries. Here we simply
take typical phenomenological profiles for Higgs-Yukawas of quark doublets ($q$) and down-type singlets ($d$) from \cite{Chankowski:2005qp}, and apply them to the leptoquark-Yukawa, leading to
\begin{align} \label{eq:U1}
\rho \sim \epsilon^2, \, \rho_d \sim \epsilon^3  ~\mbox{or} ~  \epsilon^4 \, , ~ (q) \quad  \quad \rho \sim \epsilon, \, \rho_d \sim \epsilon  ~\mbox{or} ~\epsilon^2\, ,  ~ (d)  \, ,
\end{align}
where $\epsilon$ denotes a flavor parameter of the size of the sine of the Cabibbo angle, $\epsilon \sim 0.2$.
We stress that what we mean by the hierarchies and flavor patterns in general is that they hold up to numbers (including CP phases) of order one. The overall scale $\lambda_0$ in our ansatz ($\lambda^{[\rho \kappa]}$) depends on $M$ and is fixed by $R_K$ through Eq.~(\ref{eq:RKbound}).
\footnote{
Note that here we do not explicitly  specify $U(1)_{FN }$ charges for the leptons $\ell$, whose assignments are not fixed  by phenomenology as they depend on the model of neutrino masses. 
While in general the texture Eq.\~(\ref{eq:hierarchy})  with suppressions Eq.\~(\ref{eq:U1}) is expected to hold, there is the possibility for cancellations between quark and lepton charges,  that could lead to inverted hierarchies in the leptoquark Yukawa $\lambda$. We do not consider such solutions; they are absent as long as all charges are non-negative.
}

\begin{table} [h]
\centering
\begin{tabular}{c||c|c|c}
 observable & current 90 \% CL limit  & constraint & future sensitivity \\
\hline
${\cal{B}}(\mu \to e \gamma) $ & $ 5.7 \cdot 10^{-13}$  \cite{Adam:2013mnn} & $|  \lambda_{qe}\lambda_{q\mu}^* | \lesssim   \frac{M^2}{(34 {\rm TeV})^2} $ & $6 \cdot 10^{-14}$ \cite{Baldini:2013ke}\\
${\cal{B}}(\tau \to e \gamma)$ & $1.2 \cdot 10^{-7}$ \cite{Hayasaka:2007vc} \footnote{In previous versions of this work this bound was incorrectly stated as $1.2 \cdot 10^{-8}$, which reflected the abstract of the original version of \cite{Hayasaka:2007vc} in the arXiv, which has been corrected in June 2015.} & $|  \lambda_{qe}\lambda_{q\tau}^* |  \lesssim   \frac{M^2}{(0.6 {\rm TeV})^2} $ &\\
${\cal{B}}(\tau \to \mu \gamma)$ & $4.4 \cdot 10^{-8}$   \cite{Aubert:2009ag}  &  $|  \lambda_{q\mu}\lambda_{q\tau}^* |  \lesssim   \frac{M^2}{(0.7 \, {\rm TeV})^2}$ & $5 \cdot 10^{-9}$ \cite{Aushev:2010bq}\\
${\cal{B}}(\tau \to \mu \eta)$ & $6.5 \cdot 10^{-8}$   \cite{Miyazaki:2007jp} & $|  \lambda_{s \mu}\lambda_{s \tau}^* | \lesssim   \frac{M^2}{(3.7 \,  {\rm TeV})^2}$ & $2 \cdot 10^{-9}$ \cite{Aushev:2010bq}\\
${\cal{B}}(B \to K  \mu^\pm e^\mp)$ &  $ 3.8 \cdot 10^{-8}$ \cite{Aubert:2006vb} & $\sqrt{ |  \lambda_{s \mu }\lambda_{b e}^* |^2+ |  \lambda_{b \mu }\lambda_{se}^* |^2} \lesssim   \frac{M^2}{(19.4\,  {\rm TeV})^2} $ & \\
${\cal{B}}(B \to K  \tau^\pm e^\mp) $ & $3.0 \cdot 10^{-5}$ \cite{Agashe:2014kda} & $\sqrt{ |  \lambda_{s \tau }\lambda_{b e}^* |^2+ |  \lambda_{b \tau }\lambda_{se}^* |^2} \lesssim   \frac{M^2}{(3.3 \,  {\rm TeV})^2} $ & \\
${\cal{B}}(B \to K  \mu^\pm \tau^\mp) $ & $ 4.8 \cdot 10^{-5}$  \cite{Agashe:2014kda} & $\sqrt{ |  \lambda_{s \mu }\lambda_{b \tau}^* |^2+ |  \lambda_{b \mu }\lambda_{s\tau}^* |^2} \lesssim   \frac{M^2}{(2.9 \,  {\rm TeV})^2}$ &  \\
${\cal{B}}(B \to \pi  \mu^\pm e^\mp)$ &  $ 9.2 \cdot 10^{-8}$ \cite{Aubert:2007mm} & $\sqrt{ |  \lambda_{d \mu }\lambda_{b e}^* |^2+ |  \lambda_{b \mu }\lambda_{de}^* |^2} \lesssim   \frac{M^2}{(15.6\,  {\rm TeV})^2} $ & \\
\hline
\end{tabular}
\caption{Selected LFV data, constraints and future sensitivities. Here, $q=d,s,b$. The Belle II projections  \cite{Aushev:2010bq} are for  $50 \, ab^{-1}$. 
For  the constraint from ${\cal{B}}(\tau \to  \mu \eta)$ we ignored the possibility of cancellations with $ \lambda_{d \mu}\lambda_{d \tau}^*$, see {\it e.g.,} \cite{Dorsner:2011ai}.
Following \cite{Davidson:1993qk} we ignore tuning between leading order diagrams in the $\ell \to \ell^\prime  \gamma$ amplitudes.}
\label{tab:LFV}
\end{table}

We emphasize that our main goal is to simultaneously explain lepton mixing and LNU.
Lepton mixing is structurally very different from the hierarchical quark mixing matrix, which suggests a different origin of flavor for the leptons.
We  therefore consider in Section \ref{sec:FS} scenarios where the entries between columns of $\lambda$ are related by factors that are $\pm 1$ or  $0$ as in Eq.~(\ref{eq:single}). 
Such structures  can occur naturally from a non-Abelian flavor symmetry distinguishing the generations of leptons.
We state the set of assumptions that lead to an effective factorization of quark flavor (rows) and lepton flavor (columns) in $\lambda$:
the Higgs, leptoquark and quarks are singlets under the leptonic flavor symmetries while leptons are neutral under the quark symmetries.
Most scenarios we consider will be factorized, including the  hierarchical ansatz  Eq.~(\ref{eq:hierarchy}).
Therein the flavor parameter $\kappa$ allows to split the electrons from the other leptons, and is further discussed together with LFV below. Although it is not completely general,
the structure of $\lambda^{[\rho \kappa]}$ is useful to obtain predictions, which we do in Section \ref{sec:predictions}, that can apply for more general flavor patterns, as shown in
Section \ref{sec:summaryFS}.
Non-factorized scenarios are discussed  in Section \ref{sec:A4quarks} and Appendices \ref{app:tripletDelta}, \ref{app:tdc}.

If couplings to more than one lepton flavor are present, existing constraints on LFV processes need to be considered. In Table \ref{tab:LFV} we collect LFV data, the corresponding constraints on $\lambda$-matrix elements, and give future sensitivities whenever available.
Constraints on couplings to $\tau$-leptons exist from the $B$-factory experiments Belle and Babar   \cite{Hayasaka:2007vc,Aubert:2009ag,Miyazaki:2007jp}.
These bounds are not competitive with $b$-physics ones but can be improved in the near term future at Belle II \cite{Aushev:2010bq}.
The strongest constraint on $\lambda_{q\tau}$ is therefore from $B_s$-mixing Eq.~(\ref{eq:Bsmixbound}).
Consequently, phenomenology does not currently require a split (within our approximations) between muons and taus, which is why we refrain from adding a flavor factor separating the second and the third column of $\lambda^{[\rho \kappa]}$ in Eq.~(\ref{eq:hierarchy}).
Note also that while it would be straightforward to implement such a factor, for leptoquark masses near maximal, $R_K$ already requires muon couplings $\lambda_{q\mu}$ of order one, cf.  Eq.~(\ref{eq:RKbound}), so allowing $\lambda_{q\tau}$ significantly larger than $\lambda_{q\mu}$ would challenge perturbativity. 

 The strongest constraint on LFV  stems  from $\mu \to e \gamma$ from MEG 2013
\begin{align}
|  \lambda_{qe}\lambda_{q\mu}^* | \lesssim   \frac{M^2}{(34 {\rm TeV})^2} \, , \quad q=d,s,b \, .
\end{align}
This is only mildly stronger than the $R_K$ bound Eq.~(\ref{eq:RKbound}), implying $\kappa/\rho \lesssim 0.5$.
The requirement of flavor hierarchies between electrons and muons can become more pronounced in the short-term future, with MEG upgrading their sensitivity to $6 \cdot 10^{-14}$ \cite{Baldini:2013ke}, which implies $(\kappa/\rho)_{upgrade} \lesssim 0.2$, if no signal is seen and $R_K$ remains close to its current determination.
If the $\mu \to e \gamma$ constraint tightens, it implies that in $R_K$ there has to be a dominant contribution to one lepton flavor with one branching ratio closer to its respective SM branching ratio than the other - ruling out solutions to Eq.~(\ref{eq:RKbound}) where $e$ and $\mu$ couplings are of roughly the same order of magnitude.

Current limits on LFV in $B$-decays ${\cal{B}}(B \to K  \ell^\pm \ell^{\prime \mp})$ are also given in Table \ref{tab:LFV}. They are
weaker than Eq.~(\ref{eq:RKbound}). Comparison of the bound on $B \to K  \mu^\pm e^\mp$ with the one on $B \to \pi  \mu^\pm e^\mp$, also shown, implies
$\rho_d/\rho \lesssim 1.6$.

Rare kaon decay data provide the strongest constraint on couplings to $d$-quarks  \cite{Davidson:1993qk}:
\begin{align} \label{eq:KLmm}
|  \lambda_{d \mu }\lambda_{s\mu}^* | \lesssim   \frac{M^2}{(183 {\rm TeV})^2} \, .
\end{align}
Due to chiral symmetry contributions from purely left-chiral leptons to the anomalous magnetic moment of the electron or the muons 
$\Delta a_\ell \sim |\lambda_{ql}|^2/(16 \pi^2) m_\ell^2/M^2$ are much smaller than current experimental sensitivities.

To summarize, the phenomenologically viable range for $\lambda^{[\rho \kappa]}$ parameters in Eq.~(\ref{eq:hierarchy}) is 
 \begin{eqnarray} \label{eq:param}
 \rho_d \lesssim 0.02 \, , \quad \kappa  \lesssim  0.5 \, , \quad 10^{-4} \lesssim \rho \lesssim 1  \, , \quad \kappa/\rho \lesssim  0.5   \, , \quad \rho_d / \rho \lesssim  1.6 \, . 
\end{eqnarray}
In view of $\mu \to e \gamma$, if $\rho \ll 1$ (corresponding to a low mass leptoquark), then necessarily $\kappa \ll 1$.

Regarding $\mu \to e$ conversion in nuclei, the current best limit  $\Gamma(\mu Au \to e Au)/\Gamma(\mu Au)< 7 \cdot 10^{-13}$ \cite{Bertl:2006up} is consistent with the bounds (\ref{eq:param}), however, future experiments such as COMET \cite{Cui:2009zz} and Mu2e \cite{Bartoszek:2014mya} with sensitivity below  $10^{-16}$  are sensitive to the parameter space of leptoquark models 
discussed here\footnote{We thank Franz Muheim for drawing our attention to future $\mu \to e$ conversion experiments.}. We note that corresponding tree-level contributions can always be avoided by decoupling down-quarks, suppressing $\rho_d$, or decoupling electrons, suppressing
$\kappa$.

Note that the model by \cite{Gripaios:2014tna} with the leptoquark being an $SU(2)_L$ triplet has a flavor structure 
$\lambda_{ij} \sim \epsilon^{n(qi)} \epsilon^{n(\ell j)}$, where the $\epsilon$-powers are adjusted to give the correct fermion masses, similar to what could be obtained through a FN mechanism with $U(1)$-flavor symmetries.
The resulting pattern turns out to be a viable subset of  the one presented in Eq.~(\ref{eq:hierarchy}).
Specifically,  $\rho$ and $\rho_d$ follow Eq.~(\ref{eq:U1}), and $\kappa \sim \sqrt{m_e/m_\mu}\sim 0.07$. Couplings to muons are mildly suppressed relative to the ones to taus,  of the order $\sqrt{m_\mu/m_\tau}\sim 0.24$, which is not far from our understanding of the symbol '$\sim$'  (within order one).

In Section \ref{sec:FS} we will encounter further viable patterns for $\lambda$ which are not limiting cases of either single lepton flavor Eq.~(\ref{eq:single}) or hierarchy Eq.~(\ref{eq:hierarchy}).

\section{LFV Predictions in terms of $R_K$ \label{sec:predictions}}

The leptoquark framework with the $\lambda^{[\rho \kappa]}$ ansatz from Eq.~(\ref{eq:hierarchy}) is very predictive and allows for correlations between rare $b$-decays and LFV.
 We obtain
\begin{align} 
{\cal{B}}(B \to K  \mu^\pm e^\mp) & \simeq 3 \cdot 10^{-8} \,  \kappa^2  \left( \frac{1-R_K}{0.23} \right)^2  \,, \\
{\cal{B}}(B \to K  e^\pm \tau^\mp) & \simeq  2 \cdot 10^{-8}   \, \kappa^2  \left( \frac{1-R_K}{0.23} \right)^2  \, , \\
{\cal{B}}(B \to K  \mu^\pm \tau^\mp) & \simeq 2 \cdot 10^{-8} \left( \frac{1-R_K}{0.23} \right)^2 \, ,
\end{align}
and
\begin{align}
{\cal{B}}(\mu \to  e \gamma) & \simeq 2 \cdot 10^{-12} \,  \frac{\kappa^2}{\rho^2}  \left( \frac{1-R_K}{0.23} \right)^2  \, ,\\
{\cal{B}}(\tau \to e \gamma ) & \simeq  4 \cdot 10^{-14}   \, \frac{\kappa^2}{\rho^2}  \left( \frac{1-R_K}{0.23} \right)^2  \, ,\\
{\cal{B}}(\tau \to \mu \gamma) & \simeq 3 \cdot 10^{-14} \, \frac{1}{\rho^2} \left( \frac{1-R_K}{0.23} \right)^2 \, , \\
{\cal{B}}(\tau \to \mu \eta) & \simeq 4  \cdot 10^{-11} \, \rho^2 \left( \frac{1-R_K}{0.23} \right)^2 \, .
\end{align}

For the  amplitudes of the purely leptonic decays note the  $m_\tau/m_\mu$ chiral enhancement of ${\cal{A}}(B_s \to \tau^+( \mu,e)^-)$ over
${\cal{A}}(B_s \to \tau^-  (\mu,e)^+)$ and ${\cal{A}}(B_s \to \mu^+  \mu^-)$. More  general, neglecting phase space,
\begin{align} \label{eq:ratio}
\frac{{\cal{B}}(B_s \to \ell^+ \ell^{\prime -})}{{\cal{B}}(B_s \to \ell^-  \ell^{\prime +}) }& \simeq \frac{m_\ell^2}{m_{\ell^\prime}^2}  \, .
\end{align}
This relation follows from the left-handed lepton only hypothesis, a feature of the beyond SM (BSM) models considered here,
see Appendix \ref{app:Heff} for details.
Eq.~(\ref{eq:ratio}) can be used to test  the lepton chirality in  LFV processes.

Furthermore we obtain
\begin{align}
\frac{{\cal{B}}(B_s \to \mu^+  e^-)}{{\cal{B}}(B_s \to \mu^+  \mu^-)_{\rm SM}} & \simeq   0.01 \,  \kappa^2 \cdot \left( \frac{1-R_K}{0.23} \right)^2 \, , \\
\frac{{\cal{B}}(B_s \to \tau^+  e^-)}{{\cal{B}}(B_s \to \mu^+  \mu^-)_{\rm SM}} & \simeq  4 \,  \kappa^2  \cdot \left( \frac{1-R_K}{0.23} \right)^2 \, ,\\
\frac{{\cal{B}}(B_s \to \tau^+  \mu^-)}{{\cal{B}}(B_s \to \mu^+  \mu^-)_{\rm SM}} & \simeq  4  \cdot  \left( \frac{1-R_K}{0.23} \right)^2  \, , 
\end{align}
where ${\cal{B}}(B_s \to \mu^+  \mu^-)_{\rm SM}= (3.65 \pm 0.23) \cdot 10^{-9}$ \cite{Bobeth:2013uxa}. The current bound ${\cal{B}}(B_s \to \mu^\pm e^\mp) < 1.1 \cdot 10^{-8}$ at 90 \% CL by LHCb \cite{Aaij:2013cby} is hence at least  two to three orders of magnitude away.

In addition, for  $\ell \neq \ell^\prime$,
\begin{align}
\frac{{\cal{B}}(B \to \pi   \ell^\pm \ell^{\prime \mp})}{{\cal{B}}(B \to K  \ell^\pm \ell^{\prime \mp}) } \, ,  \frac{{\cal{B}}(B_s \to K   \ell^\pm \ell^{\prime \mp})}{ {\cal{B}}(B \to K  \ell^\pm \ell^{\prime \mp})} \, , 
\frac{{\cal{B}}(B \to \ell^\pm \ell^{\prime \mp})}{{\cal{B}}(B_s \to \ell^\pm \ell^{\prime \mp})} \simeq \left( \frac{\rho_d}{\rho} \right)^2 \lesssim {\cal{O}}(2-3) \,  .
\end{align}
Precise predictions for lepton flavor conserving decays we leave for future work. We stress that the calculation of the LFV hadron decays is much less complicated
as contributions of quark loops or ($q\bar q$)-resonances coupling to the electromagnetic current are absent.

There are two extreme scenarios assuming $\lambda_0 \sim 1$:

A) $\rho=O(1)$, corresponding to high mass leptoquark in the few$\times 10$ TeV range, {\it  i.e.,} out of LHC reach. In this case
$\kappa$ can be of order one, too, implying that $B \to K  \mu^\pm e^\mp$ and $\mu \to e \gamma$ could be  just  around the corner, {\it cf.} Table \ref{tab:LFV}.
Radiative $\tau$-decays  are far away while ${\cal{B}}(B \to K  (e,\mu)^\pm \tau^\mp)$ and $ {\cal{B}}(\tau \to \mu \eta)$ are at least three orders of magnitude away from their respective current limits. ${\cal{B}}(B_s \to \tau^+ \mu^-) \sim 10^{-8}$, ${\cal{B}}(B_s \to \tau^\pm  e^\mp) \lesssim 10^{-8}$ and ${\cal{B}}(B_s \to \mu^\pm  e^\mp) \lesssim 10^{-10}$.

B) $\rho \ll 1$, corresponding to light leptoquarks  up to  few TeV in mass and $\kappa \ll 1$.
As in scenario A), $\mu \to e \gamma$ can be  just around the corner, ${\cal{B}}(B_s \to \tau^+ \mu^-) \sim 10^{-8}$ and ${\cal{B}}(B \to K  \mu^\pm \tau^\mp)$  is three orders of magnitude away from its current limit.
$B \to K  (\tau,\mu)^\pm e^\mp$, $B_s \to (\mu,\tau)^\pm  e^\mp$ and $ \tau \to \mu \eta$ are  strongly suppressed,  but $\tau \to \mu \gamma$ can be found with the next round of experiments while $\tau \to e \gamma$ remains far away.

C) $\lambda_0 \ll1$,  corresponding to light leptoquarks  up to  few TeV in mass. 
This scenario arises for instance  in the model in Appendix \ref{app:FN} where the leptoquark carries FN charge to increase powers of spurion insertions.
The LFV pattern follows from the  values of $\rho, \rho_d,\kappa$ as dictated by the flavor model.

Concerning collider searches within pattern Eq.~(\ref{eq:hierarchy}) the leptoquark decays take place into second or  third generation leptons and into third generation quarks. In scenario A),
all combinations of quarks and leptons can arise at similar level except for those involving $d$-quarks, which are necessarily suppressed.

In case of  at least two more measurements rather than upper bounds on LFV in addition to $R_K$  the parameters $\rho$ and $\kappa$  can be determined,
pinning down the flavor pattern of the leptoquark coupling matrix $\lambda$ further. It is conceivable that the latter is linked to the mechanism generating flavor for SM fermions,  {\it e.g.,} with flavor symmetries, which can be probed with the rare decay data.
In Section \ref{sec:FS} we give examples of realistic flavor symmetries that give the single lepton flavor patterns Eq.~(\ref{eq:single}), and special cases of the hierarchy pattern Eq.~(\ref{eq:hierarchy}), but also further testable  pattern.
Predictions of leptoquark coupling patterns not covered by the ansatz Eq.~(\ref{eq:hierarchy}) are given in Section \ref{sec:summaryFS}.

\section{Flavor symmetries \label{sec:FS}}

In this section we illustrate how flavor symmetries generically control the shape of the leptoquark Yukawa coupling matrix $\lambda$, and how there is in general a relation to Higgs Yukawa couplings controlled by the same symmetry. The reason for this is simply that both the leptoquark $\Delta$ and Higgs doublet (or doublets) couple to same fermions, whose generations transform as specific representations of the flavor symmetry.

We use specific models to demonstrate this, and show how to obtain some special cases for $\lambda$, including $\lambda^{[e]}$, $\lambda^{[\mu]}$ from Eq.~(\ref{eq:single}), and $\lambda^{[\rho \kappa]}$ from Eq.~(\ref{eq:hierarchy}), but also new structrures.

In terms of the quark index in $\lambda$, the structure between rows depends on what kind of symmetry transformations we assign to either the $Q_i$ generations (for the LL leptoquark) or to the RH down-type quarks $d^c_i$ (for the RL leptoquark).
In frameworks where the flavor symmetry distinguishes $Q$ and $d^c$, the LL and RL leptoquark models could be very different. Here, we focus mostly on frameworks of flavor symmetries explaining leptonic mixing, where this is not the case.
If we assume all quarks are trivial singlets of whatever non-Abelian symmetry is included, one can still assign a FN charge to quarks (under $U(1)_{FN}$) leading to Eq.~(\ref{eq:U1}) or similar.
The FN mechanism relies on  having lighter generations with larger charges in order to explain the mass hierarchies through added insertions of a $\theta$ field. We choose without loss of generality to normalize the $U(1)_{FN}$ charge of $\theta$ to $-1$. If we assume that the leptoquark $\Delta$ is neutral under the $U(1)_{FN}$, the leptoquark Yukawa couplings to lighter generations of quarks is simply suppressed by as many additional insertions of $\theta$ as the ones that appear for the respective masses, resulting in Eq.~(\ref{eq:U1}).
We relegate further considerations of the $U(1)_{FN}$ charges of $\Delta$ to Appendix \ref{app:FN}, 
in order to focus on the structure of the lepton couplings, where we consider non-Abelian flavor symmetries.

Before we delve exclusively into lepton flavor models, it is interesting to consider a SUSY framework with an underlying $SO(10)$ unified gauge group, where quark and leptons are linked. This is particularly restrictive in terms of the allowed flavor structures. At this level we will consider the flavor symmetry to be continuous ($SU(3)_F$, Section \ref{sec:SU3}).

We then investigate in some detail a SUSY framework with discrete flavor symmetry $A_{4}$ in  Sections \ref{sec:AF} and \ref{sec:AM}.
We focus on $A_4$ because it is a convenient framework to obtain the observed pattern of leptonic mixing, but also due to its relative simplicity as the smallest group with triplet representations. Appendix \ref{app:A4} contains a brief primer on $A_4$.

The reason for having SUSY in all the frameworks discussed in this section is twofold: first, SUSY keeps the gauge hierarchy problem under control, which is particularly relevant when going beyond the SM (with $SO(10)$ or flavor symmetries broken at a high scale). Second, by holomorphy it allows one to separately align the different flavor symmetry breaking directions required in the respective models. Because of SUSY, we necessarily use two $SU(2)$ doublets, $h_u$ and $h_d$. 

All the frameworks also have a $U(1)_\textrm{R}$ R-symmetry, which is spontaneously broken to its $Z_{2R}$ subgroup acting like the MSSM R-parity.
Another point that all the frameworks we consider have in common is the presence of an auxiliary Abelian flavor symmetry (either $U(1)_F$, $Z_3$ or $Z_4$). As a point of notation, in each section the charge of a superfield $\phi$ under the relevant auxiliary symmetry is denoted as $\{  \phi \}$.

\subsection{$SU(3)_F \times U(1)_F \times U(1)_\textrm{R}$ unified framework \label{sec:SU3}}

We discuss now a framework exemplified by the model presented in \cite{deMedeirosVarzielas:2005ax}. Similar models were considered in \cite{deMedeirosVarzielas:2011wx, Varzielas:2012ss}, showing how to obtain large reactor angle ($\theta_{13}$). One of the main features in the framework are that the 3 generations of each fermion flavor transform as triplets of $SU(3)$.\footnote{Due to the underlying $SO(10)$ unification, if the 3 generations of lepton $SU(2)$ doublets transform as triplets, so must the generations of other SM fermions, as each generation is unified into a $16$-plet of $SO(10)$.}

In order to build flavor symmetry invariants, we add three flavons (each with 3 generations) named $\phi_{3}^i$, $\phi_{23}^i$ and $\phi_{123}^i$ (the subscript numbers are labels, the upper indices are generation indices $i=1,2,3$, anti-triplet indices of $SU(3)$). Through the details of the superpotential of the model, the flavons acquire vacuum expectation values (VEVs), breaking the flavor symmetry in specific directions:
\begin{equation}
\label{eq:SU3VEVs}
\langle \phi_{3} \rangle = (0,0,a) \,, \quad
\langle \phi_{23} \rangle = (0, b,-b) \,, \quad
\langle \phi_{123} \rangle = (c,c,c) \,, 
\end{equation}
with $b \sim 0.20 a, c \sim 0.20 b$. This VEV hierarchy is required to be of order of the Cabibbo angle, but is also related to the hierarchy between the solar and atmospheric neutrinos \cite{deMedeirosVarzielas:2005ax} and even to the magnitude of the reactor angle $\theta_{13}$ \cite{Varzielas:2012ss}.

After $SO(10)$ is broken to a left-right symmetric GUT containing $SU(2)_L \times SU(2)_R$ (before breaking to the SM), the left-handed fermions are referred as $\psi_i$ and the conjugates of the right-handed fermions as $\psi^c_j$ (where $i,j=1,2,3$ are generation indices, triplet indices of $SU(3)$), where $\psi$ contains the 3 generations of $Q$, $L$ and $\psi^c$ contains the 3 generations of $u^c, d^c, e^c$ and $\nu^c$.
The superfields containing the electroweak doublets we denote here as $h_u$, $h_d$.

The Yukawa couplings appear from non-renormalizable terms which are generally of the type $(\phi_{3}^i \psi_i) (\phi_{3}^j \psi^c_j) h_{u,d}$ and are controlled also by an auxiliary Abelian symmetry $U(1)_F$.
The VEV directions in Eq.~(\ref{eq:SU3VEVs}) are responsible for giving hierarchical structures for all the charged fermions.\footnote{Refer to \cite{deMedeirosVarzielas:2005ax, Varzielas:2012ss} for the details related with getting viable mass ratios for the 2nd generation of fermions, such as $m_\mu / m_s \neq 1$ at the GUT scale.}

We focus now on the leptoquark Yukawa structures $\lambda$ that occur naturally when leptoquarks are added to this framework. In order to study this, we need to keep in mind the transformation properties of the SM fermions, which are all triplets under $SU(3)_F$, and the charges of the flavons under $U(1)_F$, which we denote in terms of curly brackets {\it  e.g.}
$\{ \phi_3 \} =2$, $\{ \phi_{23} \} =1$ and $\{ \phi_{123} \} =3$, while $\{ h_{u,d} \} =-4$ and the SM fermions are neutral, $\{ \psi \} = \{ \psi^c \} =0$.
To predict $\lambda$ we simply need to select how $\Delta$ transforms under $SU(3)_F \times U(1)_F$. 
Here we discuss the simplest case where the leptoquark is an $SU(3)_F$ singlet.
In Appendix \ref{app:tripletDelta} we consider leptoquark triplets under $SU(3)_F$.

Note also that, due to the underlying unification $d^c$ and $Q$ both transform equally under the flavor symmetry so there is no difference in the $\lambda$ structures corresponding to a LL or RL leptoquark,
and also that structures involving components from $\psi$ ($Q$, $L$) and $\psi^c$ ($d^c$) are symmetric due to their origin from the same $SO(10)$ multiplet.
If we were considering a unified gauge group that is not left-right symmetric then the $Q$ and $d^c$ would generally transform differently under the flavor symmetry and the structures for the RL and LL leptoquark would be different. A prime example of this is $SU(5)$, where lepton doublets $L$ belong to the same GUT multiplet as $d^c$, but $Q$ is in a different GUT multiplet together with $u^c$ and $e^c$. See Appendix  \ref{app:tdc} for some possibilities.

If $\Delta$ is an $SU(3)_F$ singlet (as $h_{u,d}$), the flavor symmetry invariants shaping $\lambda$ are $\Delta (\phi^i Q_i) (\phi^j L_j)$, which are very similar to those shaping the fermion masses. Which particular flavons couple to $\Delta$ is determined by its $U(1)_F$ charge $\{ \Delta \}$:

\begin{itemize}
\item $\{ \Delta \}=-2$ leads to $\Delta (\phi_{23}^i Q_i) (\phi_{23}^j L_j)$ which is a special limit of Eq.~(\ref{eq:hierarchy}) where $\rho=1$ and $\kappa = \rho_d=0$, given in Eq.~(\ref{eq:D-2}).
\item $\{ \Delta \}=-3$ leads to $\Delta \left( (\phi_{3}^i Q_i) (\phi_{23}^j L_j) + (\phi_{23}^i Q_i) (\phi_{3}^j L_j) \right)$.
\item $\{ \Delta \}=-4$ leads to two contributions: the dominant one $x \Delta (\phi_{3}^i Q_i) (\phi_{3}^j L_j)$ and the subleading $\Delta \left( (\phi_{23}^i Q_i) (\phi_{123}^j L_j) + (\phi_{123}^i Q_i) (\phi_{23}^j L_j) \right)$, shown together in Eq.~(\ref{eq:D-4}).
\item $\{ \Delta \}=-5$ leads to $\Delta \left( (\phi_{3}^i Q_i) (\phi_{123}^j L_j) + (\phi_{123}^i Q_i) (\phi_{3}^j L_j) \right)$.
\item $\{ \Delta \}=-6$ leads to $\Delta (\phi_{123}^i Q_i) (\phi_{123}^j L_j)$, the democratic structure given in Eq.~(\ref{eq:D-6}).
\end{itemize}
As we intend to account for $R_K$, we show only the structures with non-zero entries simultaneously in $\lambda_{se}$,  $\lambda_{be}$ or simultaneously in $\lambda_{s\mu}$,  $\lambda_{b\mu}$.

The $\lambda$ flavor structure for $\{ \Delta \}=-2$ reads
\begin{equation}
\label{eq:D-2}
\lambda^{[-2]} = \lambda_0 b^2
\left( 
\begin{array}{ccc}
0 & 0 & 0 \\
0 & 1 & -1 \\
0 & -1 & 1
\end{array}
\right) \,,
\end{equation}
which is a highly predictive limit of $\lambda^{[\rho \kappa]}$ Eq.~(\ref{eq:hierarchy}) with the same exact magnitude on the 4 non-zero entries. As there is a single coupling we can absorb the order one coefficient by redefining $\lambda_0$.

Choosing $\{ \Delta \}=-4$ gives
\begin{equation}
\label{eq:D-4}
\lambda^{[-4]} = \lambda_0 b c
\left( 
\begin{array}{ccc}
0 & 1 & -1 \\
1 & 2 & 0 \\
-1 & 0 & \frac{x a^2}{b c} - 2
\end{array}
\right) \,,
\end{equation}
which due to $a \gg b\gg c$ has $\Delta$ couple predominantly to $b \tau$.
We kept $x$ explicit to separate leading and non-leading contributions and redefine $\lambda_0$ to absorb the order one coupling of the subleading contribution, which,
although suppressed by $\sim bc/(x a^2)$ could in principle account for $R_K$. However, the simultaneous presence of the kaon constraint Eq.~(\ref{eq:KLmm}) rules this out.

The democratic pattern from $\{ \Delta \}=-6$ is
\begin{equation}
\label{eq:D-6}
\lambda^{[-6]} = \lambda_0 c^2
\left( 
\begin{array}{ccc}
1 & 1 & 1 \\
1 & 1 & 1 \\
1 & 1 & 1
\end{array}
\right) \,,
\end{equation}
and very symmetric like $\lambda^{[-2]}$, but it preserves lepton universality so it can not account for $R_K \neq 1$.

The only viable texture is therefore $\lambda^{[-2]}$. In fact the only obtainable textures if there are only these 3 flavon VEVs are the textures explored so far. $\{ \Delta \} \geq 0$ has no allowed couplings due to holomorphy, and for $\{ \Delta \} < -6$ the only possibilities repeat the existing patterns as the matrix structure is driven only be the flavons contracting with $Q_i$ and $L_j$. Even at higher order it is not possible with this field content to obtain linear combinations of the textures. The reason for this is clearer when considering a specific example, so we skip $\{ \Delta \} =-7$ as it has no allowed couplings, and take $\{ \Delta \} =-8$ where:
\begin{equation}
\Delta (\phi_{23}^g Q_g) (\phi_{23}^h L_h) (\epsilon_{ijk} \phi_{3}^i \phi_{23}^j \phi_{123}^k) \,,
\label{eq:D-8}
\end{equation}
is invariant, with the same texture as $\lambda^{[-2]}$ but with overall $U(1)_{F}$ charge $+6$ added due to $SU(3)_F$ singlet $(\epsilon_{ijk} \phi_{3}^i \phi_{23}^j \phi_{123}^k)$. Even though $U(1)_F$ would allow to swap flavon contractions to $\Delta (\phi_{23}^g Q_g) (\phi_{3}^h L_h) (\epsilon_{ijk} \phi_{23}^i \phi_{23}^j \phi_{123}^k)$ and so on, all except Eq.~(\ref{eq:D-8}) automatically vanish due to the Levi-Civita tensor.

\subsection{$A_{4}\times Z_{3}\times U(1)_{FN}\times U(1)_{\textrm{R}}$ framework \label{sec:AF}}

The first $A_4$ framework we consider is of SUSY $A_{4}\times Z_{3}\times U(1)_{FN}$ models, initially proposed in \cite{Altarelli:2005yx} and with renormalizable UV completions \cite{Varzielas:2010mp} and in particular \cite{Varzielas:2012ai} which obtains non-zero reactor angle $\theta_{13}$ in full agreement with current neutrino oscillation data.
The FN mechanism \cite{Froggatt:1978nt} is implemented separately through $U(1)_{FN}$, generating the hierarchy in the charged lepton masses without requiring small Yukawa couplings; it can easily be used to justify the quark hierarchies as described in Appendix \ref{app:FN}. SUSY's holomorphy together with the $Z_{3}$ separate the charged lepton sector and the neutrino sector.
The $A_4$ triplet flavons $\phi_l$ and $\phi_\nu$ acquire vacuum expectation values (VEVs) in special directions $(1,0,0)$ and $(1,1,1)$, respectively, shaping the leptonic mixing. The fields charged as $2$ under the R-symmetry $U(1)_{\textrm{R}}$ are the alignment fields and are responsible for giving the flavons their VEV directions. We do not discuss the details of SUSY breaking, but when it occurs $U(1)_{\textrm{R}}$ is broken to leave only a $Z_{2R}$ subgroup which distinguishes the SM fermions, i.e. it acts as R-parity. We refer to Table \ref{tab:assignment} for details about the charge assignments.

\begin{table}
  \centering
  \begin{tabular}{c||c|ccc|cc|ccc|ccc|cc|cc|ccc}
     & $\Delta$ & $Q$ & $u^c$ & $d^c$ & $\nu^{c}$ & $L$ & $e^{c}$ & $\mu^{c}$ & $\tau^{c}$  & $h_{d}$ & $h_{u}$ & $\theta$  & $\phi_{l}$ & $\phi_{\nu}$ & $\xi$  & $\xi'$ & $\phi_{l}^{0}$ & $\phi_{\nu}^{0}$ & $\xi^{0}$ \\
    \hline
    $A_{4}$ & See text &$1$ & $1$ & $1$ & $3$  & $3$  & $1$  & $1''$  & $1'$  & $1$  & $1$  & $1$  & $3$  & $3$  & $1$ & $1'$  & $3$  & $3$  & $1$ \\
    $Z_{3}$ & $2$ & $0$ & $0$ & $0$ & $2$  & $1$  & $2$  & $2$  & $2$  & $0$ & $0$ & $0$ & $0$ & $2$  & $2$ & $2$ & $0$   & $2$ & $2$  \\
$U(1)_{FN}$ & 0 & \multicolumn{3}{|c|}{See App. \ref{app:FN}} & $0$ & $0$ & $2$ & $1$ & $0$ & $0$ & $0$ & $-1$ & $0$ & $0$ & $0$ & $0$ & $0$ & $0$ &  $0$ \\  
    $U(1)_{\textrm{R}}$ & $0$ & $1$ & $1$ & $1$ & $1$ & $1$ & $1$ & $1$ & $1$ & $0$ & $0$ & $0$ & $0$ & $0$ & $0$ & $0$ & $2$ & $2$ &  $2$ \\  
    $U(1)_{\textrm{Y}}$ & $1/3$, $1/6$ & $1/6$ & $-2/3$ & $1/3$ & $0$ & $-1/2$ & $+1$ & $+1$ & $+1$ & $-1/2$ & $+1/2$ & $0$ & $0$ & $0$ & $0$ & $0$ & $0$ & $0$  & $0$ \\  
    \hline
  \end{tabular}
   \caption{Field assignment within $A_4 \times Z_3$ yielding the single lepton patterns Eq.~(\ref{eq:single}).
Leptoquark assignments are described in Section \ref{sec:AF}.
We refer to Eq.~(\ref{eq:U1}) for quark FN charges, discussed in Appendix \ref{app:FN}.}
  \label{tab:assignment}
\end{table}

We have not listed the $A_4$ representation of $\Delta$ in Table \ref{tab:assignment} as it requires a more detailed discussion.
As we have done for $SU(3)_F$, we focus on the case where the leptoquark is an $A_4$ singlet. We consider $A_4$ triplet leptoquarks in Appendix \ref{app:tripletDelta}.

For completeness, we include here very briefly the charged lepton Yukawa couplings in the superpotential
\begin{equation}
w_ L = \frac{y_e}{\Lambda^3}\theta^2 \left[ \phi_l L \right] e^c h_d + \frac{y_\mu}{\Lambda^2}\theta \left[ \phi_l L \right]' \mu^c h_d + \frac{y_\tau}{\Lambda} \left[ \phi_l L \right]'' \tau^c h_d \,,
\end{equation}
where $\Lambda$ is a scale associated with the breaking of the flavor symmetry. Coupling to the $(1,0,0)$ VEV results in a diagonal mass matrix for the charged leptons where we can identify $L_1$ with $e$, $L_2$ with $\mu$ and $L_3$ with $\tau$.

For $\Delta$, the hypercharge differs for the LL and RL leptoquark.
The Yukawa coupling to leptoquarks is associated with a renormalizable superpotential term that is either $\lambda^{ij} \left[d^c_i \Delta L_j \right]$ for $SU(2)$ doublet $\Delta$ corresponding to Eq.~(\ref{eq:L_dLQ}) - or, with a $SU(2)$ triplet $\Delta$, one would have $\lambda^{ij} \left[Q_i \Delta  L_j \right]$ as a SUSY version of  Eq.~(\ref{eq:L_tLQ}). The quark generation index is $i=d,s,b$ and the lepton generation index is  $j=e,\mu,\tau$.
We assume that all three generations of $SU(2)_L$ doublet and $SU(2)_L$ singlet quarks are trivial singlets of $A_4$. We consider $SU(2)_L$ singlet down-type quarks as an $A_4$ triplet in Appendix \ref{app:tdc} as this case arises in $A_4$ unified models \cite{deMedeirosVarzielas:2006fc, Altarelli:2008bg} with $SU(5)$.

The columns of $\lambda$ are constrained because $L$ is an $A_4$ triplet. If the leptoquark is an $A_4$ singlet, the renormalizable leptoquark superpotential terms are no longer $A_4$ invariant.
A contribution to $\lambda$ appears at leading order (LO) from a non-renormalizable term where $L$ contracts with an $A_4$ triplet flavon.
There are then three options for $\Delta$ to transform under $A_4$, and 4 non-equivalent ways to build an $A_4$ invariant:
\begin{enumerate}
\item $\Delta \sim 1$. Then $\left[\phi_l^i L_i \right] \Delta$  isolates $e$, as $\left[\langle \phi_l^i \rangle L_i \right] \propto L_1$.
\item $\Delta \sim 1''$. Then $\left[\phi_l^i L_i \right]' \Delta$  isolates $\mu$, as  $\left[\langle \phi_l^i \rangle L_i \right]' \propto L_2$.
\item $\Delta \sim 1'$. Then $\left[\phi_l^i L_i \right]'' \Delta$  isolates $\tau$, as $\left[\langle \phi_l^i \rangle L_i \right]'' \propto L_3$.
\item $\left[ \langle \phi_\nu^i \rangle L_i \right] \Delta$ couples equally to all lepton generations due to $\langle \phi_\nu \rangle$.
\end{enumerate}
By coupling to the VEV $(1,0,0)$, the different options for the leptoquark representation under $A_4$ lead to a single non-vanishing column for $\lambda$.
Options 1. and 2. correspond respectively to explicit realization of the patterns $\lambda^{[e,\mu]}$ in Eq.~(\ref{eq:single}), whereas option 3. gives SM-like $R_K$ and is disfavored by the current LHCb measurement.
Option 4., which occurs regardless of $\Delta\sim 1, 1'', 1'$, preserves lepton universality at LO but as discussed later has a myriad of LNU couplings already at NLO.

For now we postpone option 4. and take the $Z_3$ charge of $\Delta$ to be $\{ \Delta \} = 2$ (following notation used in other sections for Abelian charge). This choice means that at LO $\Delta$ couples to SM fermions only through $\phi_l$.

This is a non-trivial result that we emphasize: using the same non-Abelian flavor symmetry $A_4$ and VEVs that jointly predict viable lepton mixing with large angles, one can obtain automatically leptoquark flavor structures where LNU exists due to the isolation of a single lepton generation. This is consistent with the LNU in $B$ to $K$ decays, as shown in \cite{Hiller:2014yaa}, where the isolation of the $e$ or $\mu$ generation was merely assumed.

This isolation of lepton generation takes place at LO in a generic expansion parameter of $\langle \phi \rangle /\Lambda$, where $\phi$ represents the flavons and $\Lambda$ the scale of new physics associated with the breaking of $A_4$. This can be identified more precisely in specific UV completions \cite{Varzielas:2010mp,Varzielas:2012ai}, where it may turn out the isolation is actually exact (due to e.g. missing messengers for NLO diagrams).
Generically one may still associate next-to-leading order (NLO) effects to the presence of non-zero reactor angle $\theta_{13}$, which fixes the expansion parameter $\langle \phi \rangle /\Lambda \sim 0.2$.

With a sizable expansion parameter it is important to consider contributions beyond LO, i.e. terms featuring additional flavon insertions (of the triplet flavons or of $\xi'$, a $1'$ of $A_4$).
For the $\{ \Delta \} = 2$ case that produces lepton isolation, multiple insertions of $\phi_l$ involve the combination $[\phi_l \phi_l]_{3s}$ which gives contributions to $\lambda$ that can be reabsorbed into the LO one, due to its effective $(1,0,0)$ VEV; on the other hand both $\phi_\nu$ and $\xi'$ carry non-trivial $Z_3$ charge so they can only appear as multiples of 3 so the earliest contribution appears at NNNLO which is 
of the order $(\langle \phi \rangle /\Lambda)^3$, hence sub-percent.
We conclude that in this case, contributions beyond LO can only change the structure of $\lambda$ negligibly, and lepton isolation holds to a very good approximation.

When considering effects beyond LO it is relevant to reconsider option 4., where $\{ \Delta \} = 0$ leads to LO lepton universality due to the coupling with $\phi_\nu$.
The $Z_3$ always allows adding the neutral $\phi_l$ to this scenario, so there are NLO contributions involving $[\phi_l \phi_\nu]_{3a,3s}$. The effective VEVs of the combinations are respectively $\langle \phi_l^1 (0,  \phi_\nu^2, -\phi_\nu^3) \rangle$, $\langle \phi_l^1(2 \phi_\nu^1, -\phi_\nu^2,-\phi_\nu^3) \rangle$.
For $\{ \Delta \} = 0$, $\left[\phi_l L \right] \xi \Delta$ and $\left[\phi_l L \right] \xi' \Delta$ are also allowed by $Z_3$.
LNU indeed arises  at NLO, with 4 terms each with distinct LNU structures. Closer inspection of constraints in particular  $R_K$ Eq.~(\ref{eq:RKbound}) shows that deviations from lepton universality
at order $\sim 0.2$  in the entries $\lambda^{[\rho \kappa]}$ 
\begin{equation}
\label{eq:NLO-LNU}
\lambda^{\rm NLO} \sim \lambda_0 
\left( 
\begin{array}{ccc}
\rho_d & \rho_d & \rho_d \\
\rho & \rho & \rho \\
1 & 1 & 1
\end{array}
\right) \,,
\end{equation}
are phenomenologically  viable. We conclude that LNU at NLO gives a viable pattern for $\lambda$.

One could consider assigning quarks non-trivially under $A_4$, however, in the present framework this leads to issues in obtaining viable quark masses and mixing. Instead we will explore this option further in a different framework,  $A_4 \times Z_4$, in Section \ref{sec:A4quarks}.

\subsection{$A_4 \times Z_4 \times U(1)_\textrm{R}$ framework \label{sec:AM}}

The $A_4 \times Z_4$ SUSY framework \cite{Altarelli:2009kr} is another interesting $A_4$ framework. For its renormalizable UV completions see also \cite{Varzielas:2010mp}, and \cite{Varzielas:2012ai} for considerations regarding large $\theta_{13}$.

We start with a brief comparison with the $A_4 \times Z_3$ framework discussed in the previous section:
the first difference is that the VEV of the flavon triplet coupling to the charged leptons is now $\langle \phi_l \rangle \propto (0,1,0)$.
The second one is with respect with the FN mechanism and fields - neither $U(1)_{FN}$ nor $\theta$ are present, with the charged lepton hierarchy being due to a field $\theta'$.
\footnote{To avoid confusion with the $\xi'$ of Section \ref{sec:AF}, we renamed this field to $\theta'$. In \cite{Altarelli:2009kr, Varzielas:2010mp,Varzielas:2012ai} the same field is named $\xi'$.}
The last difference is that the sector neutral under $Z_4$ is the neutrino sector whereas  with $Z_3$, the charged lepton sector was neutral. 
This is particularly relevant due to e.g. the $\xi''$ flavon neutral under $Z_{4}$.
\footnote{In this $A_4 \times Z_4$ framework, the presence of $\xi''$ allows viable $\theta_{13}$ for a region of parameter space that is not fine-tuned, as discussed in detail in \cite{Varzielas:2012ai}.}

\begin{table}[h]
  \centering
 \begin{tabular}{c||c|ccc|cc|ccc|cc|cc|ccc|ccc}
    & $\Delta$ & $Q$ & $u^c$ & $d^c$ & $\nu^{c}$ & $L$ & $e^{c}$ & $\mu^{c}$ & $\tau^{c}$  & $h_{d}$ & $h_{u}$ & $\phi_{l}$ & $\phi_{\nu}$ & $\xi$  & $\theta'$ & $\xi''$ & $\phi_{l}^{0}$ & $\phi_{\nu}^{0}$ & $\xi^{0}$ \\
    \hline
    $A_{4}$ & See text &$1$ & $1$ & $1$ & $3$  & $3$  & $1$  & $1$  & $1$  & $1$  & $1$  & $3$  & $3$  & $1$  & $1'$ & $1''$  & $3$  & $3$  & $1$ \\
    $Z_{4}$ & $2$ & $0$ & $3$ & $0$ & $2$  & $1$  & $0$  & $1$  & $2$  & $0$ & $1$ & $1$ & $0$ & $0$  & $1$ & $0$ & $2$ & $0$ & $0$  \\
    $U(1)_{\textrm{R}}$ & $0$ & $1$ & $1$ & $1$ & $1$ & $1$ & $1$ & $1$ & $1$ & $0$ & $0$ & $0$ & $0$ & $0$ & $0$ & $0$ & $2$ & $2$ &  $2$ \\  
    $U(1)_{\textrm{Y}}$ & $1/3$, $1/6$ & $1/6$ & $-2/3$ & $1/3$ & $0$ & $-1/2$ & $+1$ & $+1$ & $+1$ & $-1/2$ & $+1/2$ & $0$ & $0$ & $0$ & $0$ & $0$ & $0$ & $0$  & $0$ \\  
    \hline
  \end{tabular}
   \caption{Field  assignment within $A_4 \times Z_4$  discussed in Section \ref{sec:AM}. It leads to patterns Eq.~(\ref{eq:hierarchy}) with normal or inverted hierarchies
regarding the lepton couplings. \label{tab:Massignment}}
\end{table}

When adding a leptoquark to this framework one can again obtain LO $\lambda$ structures like Eq.~(\ref{eq:single}) by having the $Z_4$ charge of $\Delta$ be $\{ \Delta \}=2$.
At LO we have:
\begin{enumerate}
\item $\Delta \sim 1$. Then $\left[\phi_l^i L_i \right] \Delta$  isolates $\tau$ as $\left[\langle \phi_l^i \rangle L_i \right] \propto L_3$.
\item $\Delta \sim 1''$. Then $\left[\phi_l^i L_i \right]' \Delta$  isolates $e$ as  $\left[\langle \phi_l^i \rangle L_i \right]' \propto L_1$.
\item $\Delta \sim 1'$. Then $\left[\phi_l^i L_i \right]'' \Delta$  isolates $\mu$ as $\left[\langle \phi_l^i \rangle L_i \right]'' \propto L_2$.
\end{enumerate}
They are similar to options 1.,2.,3. of Section {\ref{sec:AF}}, modified slightly to account for the (0,1,0) direction of the VEV.

More significant differences arise beyond LO, because the neutrino sector is neutral.
Given that $\{ \phi_\nu \}=0$, one is allowed to add it to the LO contribution thus constructing NLO contributions involving $[\phi_l \phi_\nu]_{3a,3s}$. This is similar to what we have seen in Section \ref{sec:AF}, although the effective VEVs of the combinations are now $\langle \phi_l^2 (-\phi_\nu^1, 0 ,\phi_\nu^3) \rangle$, $\langle \phi_l^2 (-\phi_\nu^1, 2 \phi_\nu^2,-\phi_\nu^3) \rangle$.

In order to keep the model predictive, we assume now that the $\phi_\nu$ contributions to $\lambda$ are forbidden by a partial UV completion where beyond LO terms are only allowed through singlet insertions (this is a natural consequence if $A_4$ triplet messengers are absent \cite{Varzielas:2012ai}).
In this scenario the only relevant flavon beyond LO is $\xi''$, which is also neutral under the $Z_4$. NLO contributions come from one insertion of $\xi''$ and NNLO contributions from two insertions of $\xi''$. The options for $\Delta$ are:
\begin{enumerate}
\item $\Delta \sim 1$ has LO coupling to $\tau$, $\left[\phi_l^i L_i \right]' \xi'' \Delta$ NLO coupling to electron, $\left[\phi_l^i L_i \right]'' \xi'' \xi'' \Delta$ NNLO coupling to $\mu$.
\item $\Delta \sim 1''$ has LO coupling to electron, $\left[\phi_l^i L_i \right]'' \xi'' \Delta$ NLO coupling to $\mu$, $\left[\phi_l^i L_i \right] \xi'' \xi'' \Delta$ NNLO coupling to $\tau$.
\item $\Delta \sim 1'$ has LO coupling to $\mu$, $\left[\phi_l^i L_i \right] \xi'' \Delta$ NLO coupling to $\tau$, $\left[\phi_l^i L_i \right]' \xi'' \xi'' \Delta$ NNLO coupling to electron.
\end{enumerate}
We illustrate these in matrix form, defining $\kappa'' \equiv \langle \xi'' \rangle/\Lambda \sim 0.2$:
\begin{equation}
\label{eq:AM_NLO}
\lambda^{[1]} \sim \lambda_0
\left( 
\begin{array}{ccc}
\kappa'' \rho_d & \kappa''^2 \rho_d & \rho_d \\
\kappa'' \rho & \kappa''^2 \rho & \rho \\
\kappa'' & \kappa''^2 & 1
\end{array}
\right) \,, ~
\lambda^{[1'']} \sim \lambda_0
\left(\begin{array}{ccc}
\rho_d & \kappa''\rho_d & \kappa''^2 \rho_d \\
\rho & \kappa'' \rho & \kappa''^2 \rho \\
1 & \kappa'' & \kappa''^2
\end{array} \right) \,, ~
\lambda^{[1']} \sim \lambda_0 \left(\begin{array}{ccc}
\kappa''^2 \rho_d & \rho_d & \kappa'' \rho_d \\
\kappa''^2 \rho & \rho & \kappa'' \rho \\
\kappa''^2 & 1 & \kappa''
\end{array} \right) \,,
\end{equation}
where we also parametrized the quark suppression factors as in Eq.~(\ref{eq:hierarchy}) for a more direct comparison.
The first pattern,  $\lambda^{[1]}$ cannot simultaneously accommodate $R_K$ and $B_s$-mixing constraints unless $M \lesssim \mbox{few}$ TeV. The second one, corresponding to an inverted hierarchy with leptons flavor ordering of leptoquark couplings in opposite way as the one to the Higgs, and the third one, normal hierarchy (for electrons) are both viable. Phenomenological implications are summarized in Section \ref{sec:summaryFS}.

To conclude, within $A_4 \times Z_4$  it is possible to obtain special versions of Eq.~(\ref{eq:hierarchy}).
Note also that, if instead of $\xi''$ we had considered a $Z_4$ neutral $1'$ flavon - which would be difficult to distinguish in terms of lepton mixing angles \cite{Varzielas:2012ai} - for the same LO contribution, replacing the 1'' with the 1' leads to swapping which lepton generation is coupling at NLO and at NNLO, {\it i.e}, effectively swapping
in each matrix in Eq.~(\ref{eq:AM_NLO}) the $\kappa''$ with the $\kappa''^2$ terms. The resulting
$\lambda$ structures are different and can be tested experimentally.
As a corollary of that, by having simultaneously both $1'$ and $1''$ one can obtain LO in one lepton generation and at NLO the other two, {\it e.g.} LO coupling to $\mu$, NLO to $e$ and to $\tau$. The variant with LO coupling to $\tau$ is viable only for leptoquark masses not exceeding a few TeV due to Eq.~(\ref{eq:RKbound}) and Eq.~(\ref{eq:Bsmixbound}).

\subsubsection{Quarks non-trivial under $A_4$ \label{sec:A4quarks}}

An interesting option that can be considered is to make different generations of quarks  transform as different non-trivial singlets of $A_4$. Consider option 1., 2., 3. of the $A_4 \times Z_4$ framework.
If all quark generations are the same (non-trivial) $A_4$ singlet, it just shifts which lepton generation is isolated at LO by each leptoquark choice. For instance, if $Q_i$ are all $1'$, then $\Delta \sim 1''$ is the leptoquark that would now couple to $\tau$, whereas $\Delta \sim 1$ which previously isolated $\tau$ would now couple instead to $\mu$.

On the other hand new flavor pattern arise when different generations of quarks are transforming under different $A_4$ singlets. We note that some care is necessary as this possibility may lead to unwanted implications for the Yukawa couplings with the Higgs and a CKM matrix that is not viable. Indeed this is generally the case for the $A_4 \times Z_3$ framework.

What happens with different generations of quarks assigned as different $A_4$ singlets is that, depending on the quark generation, the leptoquark couples to different columns at each order.
As an illustration of this, and neglecting the quark FN charges for simplicity,
take $\Delta \sim 1'$, $d^c_{2,3}$ still as trivial singlets but $d^c_{1} \sim 1''$. We have at LO:
\begin{equation}
\Delta \left[ \phi_l^i L_i \right]'' \left(x_2  d^c_2  + x_3  d^c_3 \right) + \Delta \left[ \phi_l^i L_i \right] (x_1 d^c_1)
\end{equation}
corresponding to:
\begin{equation}
\label{eq:A4quarks}
\lambda^{LO} = \lambda_0
\left( 
\begin{array}{ccc}
0 & 0 & x_1 \\
0 & x_2 & 0 \\
0 & x_3 & 0
\end{array}
\right) \,.
\end{equation}
Such patterns generalize lepton flavor isolation Eq.~(\ref{eq:single}), and at the same time are no longer special limits of  Eq.~(\ref{eq:hierarchy}).
The pattern Eq.~(\ref{eq:A4quarks}) could be distinguished from the others for instance through the tree-level decays of the leptoquark in colliders.

Viable masses for quarks can be obtained for this type of $d^c$ assignments under $A_4$  as long as there are flavons allowing invariants populating the respective columns of the Yukawa coupling to the Higgs, $h_d Q_i Y^d_{ij} d^c_j$. This is not the case for the $A_4 \times Z_3$ framework, but in the $A_4 \times Z_4$ framework considered in this section the $Z_4$ neutral flavon allows it. For the example considered above where $d^c_{1} \sim 1''$, terms populating $Y^d_{ij}$ with $j=1$ can be obtained through a double insertion of the flavon $\xi''$ (note that if there is a FN charge of $d^c_1$ it should be amended with the additional flavon insertions in mind cf. Appendix \ref{app:FN}). For this reason, choices leading to $Q_3 u^c_3$ non-trivial under $A_4$ are unappealing from a theoretical standpoint, as having the top quark mass arise from a renormalizable term is preferable.

In general, constructions of this type yield patterns for $\lambda^{LO}$ where a single quark flavor can couple to a single lepton flavor. Due to $R_K$,
either electrons or muons have to couple to both $s$ and $b$. Therefore,
the following structures in addition to Eq.~(\ref{eq:A4quarks}) are phenomenologically viable:
\begin{equation}
\label{eq:A4quarksmore}
\left( 
\begin{array}{ccc}
x1 & 0 & 0 \\
0 & x_2 & 0 \\
0 & x_3 & 0
\end{array}
\right) \,, \quad   \quad 
\left( 
\begin{array}{ccc}
0 & 0 & x_1 \\
 x_2 & 0 & 0 \\
 x_3 & 0& 0 
\end{array}
\right) \,,
\quad  \quad 
\left( 
\begin{array}{ccc}
0 & x_1 & 0 \\
x_2 & 0 & 0 \\
 x_3 & 0 & 0 
\end{array}
\right) \,.
\end{equation}
All of these effective two-flavor patterns predict LFV involving two lepton flavors only at LO, but receive respective NLO (and NNLO) contributions which are similarly shifted in the first row and are also themselves a two-flavor pattern.

However, in specific UV completions it is possible to forbid NLO and NNLO contributions to $\lambda$, while allowing the desired LO contributions and the necessary $Y^d_{ij}$ Yukawa couplings.  \footnote{A relatively simple possibility for RL leptoquarks is to forbid  $A_4$ triplet and $SU(2)$ singlet  messengers. We have checked in this case the absence of pattern-changing beyond-LO diagrams, while the required diagrams for RL leptoquark Yukawa couplings and the $h_d$ Yukawa to quarks and charged leptons are mediated by $SU(2)$ doublets.}
In fact, allowing only a minimal set of messengers we can obtain $\lambda$ matrices like Eq.~(\ref{eq:A4quarks}) with $x_1=0$, corresponding to the truly minimal patterns  from \cite{Hiller:2014yaa} that  explain LNU data Eq.~(\ref{eq:RKdata}).

\subsection{Summary of flavor symmetry frameworks and predictions \label{sec:summaryFS}}

We summarize in Table \ref{tab:sum_lambda} the features of the $SU(3)_F$ and $A_4$ frameworks discussed in the previous sections, for a single leptoquark.

\begin{table} [h]
\centering
\begin{tabular}{c||c|c|c}
$\lambda$ structure & symmetry & flavons & $\Delta$ assignment\\ \hline
Eq.~(\ref{eq:hierarchy}) with zeros, i.e.~Eq.~(\ref{eq:D-2}), $\lambda^{[-2]}$ & $SU(3)_F \times U(1)_F$ & $\langle \phi_{23} \rangle = (0,b,-b)$ & $\{ \Delta \}=-2$ \\
Eq.~(\ref{eq:single}), $\lambda^{[e]}$ electrons only& $A_4 \times Z_3$ & $\langle \phi_l \rangle = (u,0,0)$ & 1,  $\{ \Delta \}=2$\\
Eq.~(\ref{eq:single}), $\lambda^{[\mu]}$ muons only & $A_4 \times Z_3$ & $\langle \phi_l \rangle = (u,0,0)$ & 1'',  $\{ \Delta \}=2$\\
Eq.~(\ref{eq:hierarchy}) with $\kappa\sim 1$, i.e. Eq.~(\ref{eq:NLO-LNU}), $\lambda^{\rm NLO}$& $A_4 \times Z_3$ & $\langle \phi_\nu \rangle = (w,w,w)$ & $1^x$,  $\{ \Delta \}=0$\\
Eq.~(\ref{eq:hierarchy}), normal hierarchy, i.e.~Eq.~(\ref{eq:AM_NLO})$\lambda^{[1']}$ & $A_4 \times Z_4$ & $\langle \phi_l \rangle = (0,u,0)$, $\xi''$ & 1',  $\{ \Delta \}=2$\\
Eq.~(\ref{eq:hierarchy}), inverted hierarchy, i.e.~Eq.~(\ref{eq:AM_NLO})$\lambda^{[1'']}$ & $A_4 \times Z_4$ & $\langle \phi_l \rangle = (0,u,0)$, $\xi''$ & 1'',  $\{ \Delta \}=2$\\
Eq.~(\ref{eq:A4quarks}) , Eq.~(\ref{eq:A4quarksmore}) two-flavor & $A_4 \times Z_4$ &  $\langle \phi_l \rangle = (0,u,0)$  & 1' ,  $\{ \Delta \}=2 $ \\
\hline
\end{tabular}
\caption{Summary of viable $\lambda$ structures and how they can be obtained through a flavor symmetry. In all cases except
Eq.~(\ref{eq:D-2}) and Eq.~(\ref{eq:A4quarks}) the quarks transform trivially under the  flavor symmetry.}
\label{tab:sum_lambda}
\end{table}

The phenomenology of the patterns following from the hierarchy ansatz Eq.~(\ref{eq:hierarchy}) has been detailed in Section \ref{sec:predictions}.
The two-lepton flavor patterns Eq.~(\ref{eq:A4quarks}) and Eq.~(\ref{eq:A4quarksmore}) discussed in Section \ref{sec:A4quarks} that arose newly in the analysis of the flavor symmetries, have predictions similar to the
single lepton flavor pattern Eq.~(\ref{eq:single}): they successfully explain $R_K$ by either BSM in $b\to s e^+e^-$ or $b\to s \mu^+ \mu^-$ transitions, however LFV signals
can appear in addition in either $e \mu$, $e \tau$ or $\mu \tau$ related to $b\to d$ or $s \to d$ transitions. For the example
in Eq.~(\ref{eq:A4quarks}): $B \to \pi \tau \mu$, $B \to \tau \mu$ and $\tau \to K^{(*)} \mu$,  but none involving electrons.
The sub-case for $x_1=0$, that is no couplings to $d$-quarks, can also be obtained, corresponding to the minimal benchmark patterns presented in \cite{Hiller:2014yaa}, to which we refer for their phenomenology. They have no LFV. Implications include effects in $b \to (d,s) \nu \bar \nu$ decays, relevant for the Belle II experiment. 
Correlations between $B \to K \nu \bar \nu$ and $B \to K^* \nu \bar \nu$ have been worked out  in \cite{Buras:2014fpa}.

The other genuine non-Abelian type of $\lambda$ structure is the inverted hierarchy Eq.~(\ref{eq:AM_NLO}), $\lambda^{[1'']}$, in which the leptoquarks couplings do not follow the ones of the leptons to the Higgs. This pattern predicts LFV  in processes involving $e \mu$, $e \tau$ and $\mu \tau$, which can be read off from the
estimates given in Section \ref{sec:predictions} using $\kappa =\kappa^{\prime \prime}$,  $\kappa =\kappa^{\prime \prime 2}$ and $
\kappa =\kappa^{\prime \prime 3}$, respectively. The normal hierarchy pattern,  $\lambda^{[1']}$ of Eq.~(\ref{eq:AM_NLO}), predicts LFV  in processes involving $e \mu$, $e \tau$ and $\mu \tau$ corresponding to $\kappa =\kappa^{\prime \prime 2}$,  $\kappa =\kappa^{\prime \prime 3}$ and $
\kappa =\kappa^{\prime \prime }$, respectively. In these cases, $\kappa^{\prime \prime} = {\cal{O}}(\theta_{13})$.

While it appears that it is possible to obtain in each flavor model pattern for RL and LL leptoquarks alike, we stress that this is a feature of the specific frameworks considered where $Q$ and $d^c$ transform equally under the flavor symmetry, either due to underlying unification or for the sake of simplicity.

\section{Rare Higgs decays \label{sec:higgs}}

In this section we discuss leptoquark effects in decays of the Higgs into fermion anti-fermion.
Such decays have received recent interest with the advent of such a particle and its various decay modes being stringent test of the flavor sector of the SM and beyond,
{\it e.g.}, recently, \cite{Harnik:2012pb,Dery:2014kxa}.

\begin{figure}[h]
\centering
\includegraphics[width=6 cm]{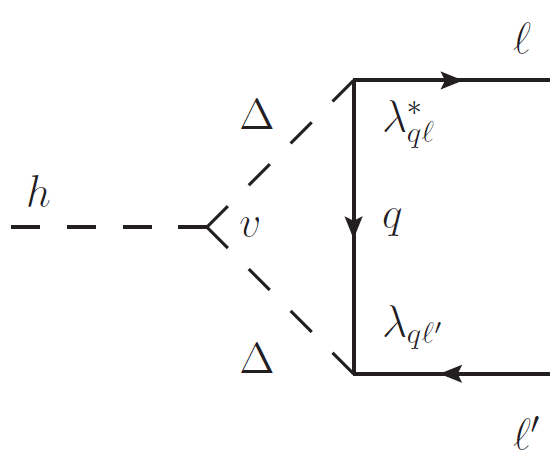}
\caption{1-loop contribution to $h \to \ell \ell'$ from $\Delta$.}
\label{fig:LQloop}
\end{figure}

Leptoquark contributions to  Higgs decays are induced at one-loop as exemplified in Figure \ref{fig:LQloop}. 
The amplitude is proportional to a renormalizable term
$\Delta^\dagger \Delta h^\dagger h$, whose coefficient in general is model-dependent. After electroweak symmetry breaking a coupling of the Higgs to two leptoquarks is induced
at order $v h \Delta \Delta^*$, where $v\simeq 174$ GeV sets the electroweak scale. The corrections to flavor-diagonal modes $h \to \ell \ell$ are hence parametrically given as
\begin{align} \label{eq:Higgscorrections}
\frac{\delta y_{ \ell}}{y_\ell} \sim N_c \frac{v^2}{M^2} \frac{| \lambda_{q \ell } |^2}{16 \pi^2} \, , \quad q=d,s,b \, ,
\end{align}
where $N_c=3$ denotes the number of colors.
Analogously for $h \to qq$:
\begin{align} \label{eq:Higgsqcorrections}
\frac{\delta y_{q}}{y_q} \sim \frac{v^2}{M^2} \frac{| \lambda_{q \ell } |^2}{16 \pi^2} \, , \quad \ell=e,\mu,\tau \, . 
\end{align}
Contributions from further  one-loop diagrams involving the Higgs-Yukawa couplings exist but are suppressed further by at least $y_{b,\tau} \ll 1$, and contributions with $y_t$ don't bring in additional enhancements either.
The relative corrections $\delta y/y$ to the diagonal Higgs Yukawas therefore do not exceed the $10^{-2}$ level.

Spurion analysis shows when LFV in the Higgs coupling $\bar \ell_L X e_R$ is induced, {\it cf.} Eq.~(\ref{eq:L_dLQ}),
\begin{align} 
X=A Y_\ell + B \lambda^\dagger \lambda Y_\ell + \ldots \, ,
\end{align}
hence the leptoquark couplings, in general, induce LFV couplings. A single Higgs doublet as in the SM suffices.
Note that single lepton-flavor patterns as in Eq.~(\ref{eq:single}) fail to induce $X_{i j}$ for $i\neq j$.

Leptoquark contributions to the $B$-term in the spurion expansion arise at order 
\begin{align} \label{eq:HiggsLFV}
\frac{y_{ \ell \ell^\prime }}{y_{\ell^\prime}} \sim N_c \frac{v^2 }{M^2} \frac{ \lambda_{q \ell }^* \lambda_{q \ell^\prime}}{16 \pi^2} \, , \quad q=d,s,b \, .
\end{align}
$y_{  \ell^\prime \ell }$ is suppressed relative to $y_{  \ell \ell^\prime }$ by $y_\ell/y_{\ell^\prime}$ for $m_\ell < m_{\ell^\prime}$.

The coupling in Eq.~(\ref{eq:HiggsLFV}) is the same which drives $\ell^\prime \to \ell \gamma$.
We find that, using the limits from Table \ref{tab:LFV},  LFV Higgs effects are limited as
\begin{align} \label{eq:LFV-LQ}
\frac{y_{ \mu \tau}}{y_\tau} \lesssim 10^{-3} \, , \quad \frac{y_{e  \tau}}{y_\tau} \lesssim 10^{-3} \, , \quad \frac{y_{e  \mu}}{y_\mu} \lesssim 10^{-6} \, .
\end{align}
Flavor models of course predict patterns for these couplings and relate $h \to \ell^\prime \ell$ decays. {\it E.g.,} in terms of the parametrization Eq.~(\ref{eq:hierarchy}):
\begin{align}
\frac{ y_{ e \tau }}{ y_{ \mu \tau }}\sim \kappa  \, , \quad \quad  \frac{ y_{ e \mu}}{ y_{ \mu \tau }}\sim \kappa  \, .
\end{align}
If $\kappa$ is large, this implies that all LFV branching ratios are strongly suppressed due to the suppression of $h \to \mu e$.
If $\kappa$ is small,  the $h \to \tau \mu$ branching ratio is much larger than the ones involving electrons.
Different patterns follow from other flavor structures of $\lambda$ summarized in Section \ref{sec:summaryFS}.

The CMS data on the $h \to \tau \mu$ branching fraction \cite{Khachatryan:2015kon} imply a sizable off-diagonal Yukawa-coupling to the SM-like Higgs 
$\sqrt{|y_{\tau\mu}|^2+|y_{\mu \tau}|^2}=(2.6 \pm 0.6) \cdot 10^{-3}$,
about a third of the coupling to the taus itself, $ y_\tau \simeq 0.01 \tan \beta$. Such a sizable effect, taken at face value, exceeds the leptoquark  estimates Eq.~(\ref{eq:LFV-LQ}) and points to mechanisms beyond perturbative loops.
An example is given by \cite{Crivellin:2015mga,Crivellin:2015lwa}.

\section{Conclusions \label{sec:conclusion}}

The bottom-up BSM scenario with leptoquarks considered here provides further evidence of how new physics can help to learn about flavor \cite{Nir:2007xn}: if the anomalous signatures in the FCNCs Eq.~(\ref{eq:RKdata}) hold
they offer unique possibilities to probe the mechanism of flavor, such as the type and charge assignment of a flavor symmetry whose imprints on SM matter were the only experimental information previously available.

By adding a leptoquark to existing flavor symmetry frameworks we obtained sample scenarios  where the flavor symmetries are simultaneously responsible for LNU ($R_K$) and LFV, drawing in each case relations to the lepton mixing angles and charged fermion hierarchies predicted in these models.

The predictions for LFV  lepton and $b$-decays are copious, and we summarize them in Sections \ref{sec:predictions} and \ref{sec:summaryFS}. The  LFV branching ratios, whose
size is driven by $(1-R_K)^2$ are in part
accessible to near-term future  experiments, including MEG, LHCb and Belle II,  the details of which depend on the specific flavor pattern of the leptoquark coupling $\lambda$,
single lepton flavor Eq.~(\ref{eq:single}), the hierarchy pattern Eq.~(\ref{eq:hierarchy}), or further patterns which follow from the non-Abelian nature of the underlying  flavor symmetry.
Viable and realistic ones are summarized in Table \ref{tab:sum_lambda}. They can be distinguished phenomenologically. Further data could not only be a discovery of BSM, but also provide clues about
flavor. Already current data rules out many choices of flavor group assignments, as demonstrated in Section \ref{sec:FS}.

Indeed if certain patterns of leptoquark coupling are observed in the future, they constitute strong hints of the presence of non-Abelian flavor symmetries. From the examples studied in this paper we single out three patterns. One is predicted in the unified $SU(3)_F$ framework, and is  highly symmetrical with four entries of the same magnitude, Eq.~(\ref{eq:D-2}).
Alternatively if lepton isolation, either in electrons on muons, Eq.~(\ref{eq:single}), is observed to below percent level, that points  to models such as our $A_4 \times Z_3$ which only allow deviations from isolation at the NNNLO - or, given an appropriate UV completion, to models like our $A_4 \times Z_4$.
Mimicking these patterns without using non-Abelian symmetries would require rather unnatural leptoquark couplings.

If LNU and LFV in rare $b$-decays is observed, it allows to completely identify the chirality-structure of the leptoquark-lepton-quark couplings:
comparison of $R_K$ with related non-universality tests into others strange final states, such as $K^*, X_s, ...$ allows to probe for 
right-handed quark currents \cite{Hiller:2014ula}.
Specifically, the models Eq.~(\ref{eq:L_dLQ})  and  Eq.~(\ref{eq:L_tLQ}) can be distinguished. Pinning down the chirality of the leptons is possible by comparing
the LFV branching ratios $B_{(s)} \to \ell^+ \ell^{\prime -}$ with $B_{(s)}  \to \ell^- \ell^{\prime +}$, which pick up different lepton mass factors.
This method of diagnosing quark and lepton chirality, of course, applies model-independently for the low energy theory Eq.~(\ref{eq:Heff}), {\it i.e.,} is not limited to leptoquark mediated FCNCs.

The current hint for BSM in  $|\Delta B|=1$ transitions in $R_K$ provides an anchor for fixing BSM scales when interpreted as a ``signal'' and combined  with $|\Delta B|=2$ ($B_s$-mixing) bounds. The mass range of the leptoquarks is determined to be
right above search limits around ${\cal{O}}(\mbox{TeV})$ and  below ${\cal{O}}(50)$TeV \cite{Hiller:2014yaa}.  In $pp$ collisions the leptoquarks will be pair-produced 
and decay  in our frameworks predominantly into second or  third generation leptons and into third generation quarks. 

{\bf Note added:}
During the completion of this work a paper \cite{Dorsner:2015mja}, appeared, where related leptoquark effects to $h \to \ell \ell'$ were considered. 

\section*{Acknowledgements}

This project is supported by the European Union's
Seventh Framework Programme for research, technological development and
demonstration under grant agreement no PIEF-GA-2012-327195 SIFT (IdMV) and
by the DFG Research Unit FOR 1873 ``Quark Flavour
Physics and Effective Field Theories'' (GH).
The authors would like to thank the organizers of the NExT workshop 2014 "News from BSM, Higgs and Supersymmetry" at U. of Sussex, where this project was initiated.

\appendix

\section{Leptoquarks in $b$-decays \label{app:Heff}}

This appendix explains how to obtain predictions for $b$-decays from the leptoquark interactions  Eq.~(\ref{eq:L_dLQ}) and Eq.~(\ref{eq:L_tLQ}), corresponding 
to $\Delta(3,2)_{1/6}$  and $\Delta(3,3)_{-1/3}$, respectively. After integrating-out and fierzing \cite{Davidson:1993qk}, $b \to s \ell^+ \ell^{\prime -}$ transitions with  
$\ell =\ell^\prime$ and
$\ell \neq \ell^\prime$ are induced.
Employing the common
$|\Delta B|=|\Delta S|=1$ effective Hamiltonian
\begin{align} \label{eq:Heff}
{\cal{H}}_{\rm eff} =- 4\frac{G_F}{\sqrt{2}} V_{tb} V_{ts}^* \frac{\alpha_e}{4 \pi }\sum_i C_i O_i \, ,
\end{align}
where $\alpha_e$, $V_{ij}$ and $G_F$ denote the fine structure constant, the CKM matrix elements and Fermi's constant, respectively,
the semileptonic operators 
\begin{align} 
  {\cal{O}}_{9} & =  \bar{s} \gamma_\mu P_{L} b \, \bar{\ell^\prime} \gamma^\mu \ell \,, \quad
  {\cal{O}}_{10}  = \bar{s} \gamma_\mu P_{L} b  \, \bar{\ell^\prime} \gamma^\mu \gamma_5 \ell \,, \\
   {\cal{O}}_{9}^\prime & =  \bar{s} \gamma_\mu P_{R} b \, \bar{\ell^\prime} \gamma^\mu \ell \,, \quad
  {\cal{O}}_{10}^\prime  = \bar{s} \gamma_\mu P_{R} b  \, \bar{\ell^\prime} \gamma^\mu \gamma_5 \ell \,, 
\end{align}
receive leptoquark contributions as \cite{Hiller:2014yaa} 
\begin{align} \nonumber
C_9 &=-C_{10} =\frac{\pi}{\alpha_{e}} \frac{\lambda_{s\ell^\prime}^*\lambda_{b\ell}}{V_{tb} V_{ts}^*}
\frac{\sqrt{2}}{2M^2G_F}  \, \quad  \mbox{for} ~~\Delta(3,3)_{-1/3} \, , \\
C_{10}' &=-C_9'=\frac{\pi}{\alpha_{e}}
    \frac{\lambda_{s \ell}\lambda_{b \ell^\prime}^*}{V_{tb}V_{ts}^*}\frac{\sqrt{2}}{4 M^2 G_F}    \quad  \quad  \mbox{for} ~~\Delta(3,2)_{1/6} \, . \label{eq:left}
\end{align}
Note that $C_9^{(\prime)}+C_{10}^{(\prime)}=0$ shows  that the leptons are left-handed.

The amplitude for $B_s \to \ell^+ \ell^{\prime -}$ decays is calculated as
\begin{align} 
\langle 0| \bar s \gamma_\mu P_{L/R} b| B_s \rangle \cdot  \bar u(\ell^{\prime -}) \left[ C_9^{(\prime)}  \gamma^\mu + C_{10}^{(\prime)} \gamma^\mu \gamma_5 \right] v(\ell^+) 
\propto m_{\ell^\prime}(C_{10}^{(\prime)} +C_9^{(\prime)})  + m_\ell (C_{10}^{(\prime)} -C_9^{(\prime)}) 
\, , \label{eq:amp}
\end{align}
where  we used that the  hadronic matrix element is proportional to the $B_s$ mesons' four-momentum, which equals the 
sum of the leptons' four-momenta $q_\mu$, and then applied the equations of motion for particle  $u$ and anti-particle spinors $v$. For same flavor leptons therefore the vector current contribution vanishes and
using Eq.~(\ref{eq:left}) the r.h.s. of  Eq.~(\ref{eq:amp}) reads $2 m_\ell C_{10}^{(\prime)}=-2 m_\ell C_9^{(\prime)} $, with proportionality given by the anti-particle mass.
Right-handed leptons, corresponding to $C_9^{(\prime)}-C_{10}^{(\prime)}=0$ would give  instead the mass of the
negatively charged lepton as chiral factor.

The model-independent framework Eq.~(\ref{eq:Heff}), Eq.~(\ref{eq:left}) allows to compute leptoquark effects in $B \to K^{(*)} \mu \mu$ decays in a straightforward way. Correlations of the global fits and $R_K$ in such models have been discussed recently in \cite{Hiller:2014yaa,Gripaios:2014tna}. 
A detailed exploration including the very recent, preliminary $3 {\rm fb}^{-1}$ data by LHCb on
$B \to K^* \mu \mu$ angular observables \cite{LHCb:2015dla} taking into account SM uncertainties is beyond the scope of our work.

\section{Quarks and Froggatt-Nielsen \label{app:FN}}

Relative suppression between rows of $\lambda$ is naturally obtained by assigning different $U(1)_{FN}$ charges to the generations of  quarks. The outcome depends also on the charge of the leptoquark. In this appendix curly brackets denote the $U(1)_{FN}$ charge of a field. 

Explicitly, take the leptoquark $SU(2)$ triplet, $\{ \Delta \}=0$ and $\{ Q_{1,2,3} \}=3,2,0$   respectively \cite{Chankowski:2005qp}.
The leptoquark couplings would be:
\begin{equation}
\label{tLQnonren}
 \left(x_d Q_1 \Delta  \epsilon^3 + x_s   Q_2 \Delta \epsilon^2 + x_b Q_3 \Delta \right) \ell \, ,
\end{equation}
where $\epsilon = \langle \theta \rangle / \Lambda \sim 0.2$.
$x_{d,s,b}$ are order one dimensionless couplings.
The outcome is an hierarchy between rows of the leptoquark couplings
$\lambda_{d \ell}  < \lambda_{s \ell} < \lambda_{b \ell}$ as e.g. $\lambda_{d \ell}/ \lambda_{s \ell}= \frac{x_d}{x_s} \epsilon$, irrespective of lepton flavor.
One can easily generalize this for alternative assignments of FN charges, such as $\{ Q_{1,2,3} \}=4,2,0$ \cite{Chankowski:2005qp}, and similarly for the leptoquark  $SU(2)$ doublet, where  $\{ d^c_{1,2,3} \}=2,1,0$ or $\{ d^c_{1,2,3} \}=1,1,0$  \cite{Chankowski:2005qp}. Values for suppression factors of strange quarks versus $b$-quarks, $\rho$, and  down quarks versus $b$-quarks, $\rho_d$,
can be read-off and are compiled in Eq.~(\ref{eq:U1}). In this framework $\rho_d/\rho \lesssim 1$, consistent with the bounds Eq.~(\ref{eq:param}).

Choosing non-zero values of FN-charges for the leptoquark significantly impacts $\lambda$ and the resulting phenomenology, which forces $\{ \Delta \}$ to be within 0 and $\sim 3$: $\{ \Delta \} > 0$ adds an overall suppression to all leptoquark couplings of $\epsilon^{\{ \Delta \}}$, which, while keeping $\rho_d/\rho$ unchanged, leads to suppression of  $\lambda_0$. This requires lighter leptoquarks to explain $R_K$. This case is discussed in scenario C of Section \ref{sec:predictions}. Too large charges $\{ \Delta \} \gtrsim 3$ cannot accommodate simultaneously Eq.~(\ref{eq:RKbound}) and direct search limits $M \gtrsim 1$ TeV.
Within SUSY negative charges are in conflict with  Eq.~(\ref{eq:RKbound}), because due to holomorphy, $\{ \Delta \} < 0$ forbids couplings to $b$-quarks.
To see this take $\{ \Delta \} = -1$ and compare to $\{Q_3\}$ or $\{d_3^c\}$, which are zero to account for the mass of the $b$ and top quarks.

\section{A few $A_4$ details \label{app:A4}}

$A_4$ has 4 distinct irreducible representations, three of them are 1-dimensional, i.e. singlets, and one of them is 3-dimensional, i.e. the $A_4$ triplet.
The trivial singlet we label $1$, and it transforms trivially under $A_4$.
The (non-trivial) singlets $1'$ and $1''$ are conjugate to one another and
they transform under a specific $A_4$ generator, $T$, by getting multiplied respectively by $\omega^2$ and $\omega$ ($\omega\equiv e^{i 2 \pi/3}$, with $\omega^3 \equiv 1$).
The product of $A_4$ singlets has $(1 \times 1')$, $(1 \times 1'')$, $(1' \times 1'')$ transforming as $1'$, $1''$, and $1$ respectively.
The action of the group on triplets is represented by $3\times 3$ matrices. Consider specific triplets $A=(a_1,a_2,a_3)$, $B=(b_1,b_2,b_3)$. Under generator $T$, which is diagonal in the basis we are considering, we have $T A = (a_1, \omega^2 a_2, \omega a_3)$ (same for $B$).
We use the conventions in \cite{Varzielas:2010mp, Varzielas:2012ai}, and square brackets to indicate $A_{4}$ products:
\begin{align}
\left[ A B \right]= &(a_1 b_1 + a_2 b_3 + a_3 b_2) \sim 1 \, , \\
\left[ A B \right]' = &(a_1 b_2 + a_2 b_1 + a_3 b_3) \sim 1'  \, ,\\
\left[ A B \right]'' = &(a_1 b_3 + a_2 b_2 + a_3 b_1) \sim 1'' \,.
\end{align}
It is also possible to construct a symmetric ($s$) and an anti-symmetric ($a$) triplet:
\begin{equation}
        [AB]_s =\frac{1}{3}\left(\begin{array}{c}
                                     2a_1 b_1-a_2 b_3-a_3 b_2\\
                                     2a_3 b_3-a_1 b_2-a_2 b_1\\
                                     2a_2 b_2-a_3 b_1-a_1 b_3\\
                                     \end{array}
                               \right) \, , \quad
        [AB]_a=\frac{1}{2}\left(\begin{array}{c}
                                 a_2 b_3-a_3 b_2\\
                                 a_1 b_2-a_2 b_1\\
                                 a_3 b_1-a_1 b_3\\
                                \end{array}\right) \, .
\end{equation}

\section{(Anti-)triplet $\Delta$ \label{app:tripletDelta}}

Within the frameworks we considered in Section \ref{sec:FS} it also possible to consider 3 generations of leptoquarks transforming as a representation of the flavor symmetry.
This is more intricate than the single leptoquark scenarios we focused on, as there are three $\lambda$ structures (one for each generation of the $\Delta$ multiplet), and the flavor structure of the mass matrix for the $\Delta$ generations is also relevant. In the frameworks considered here $\Delta$ only acquire masses after the R-symmetry is broken (the leptoquark mass terms behave similarly to the $\mu$-term $\mu h_u h_d$). As we will show, in some cases there are holomorphic $\Delta$ bi-linears that could couple to whatever superfield is responsible for breaking the R-symmetry.

\subsection{$SU(3)_F$}

In the $SU(3)_F$ framework one can have anti-triplet $\Delta^i$ (i.e. 3 generations of leptoquark transforming like the $\phi^i$) or instead a triplet $\Delta_i$ (i.e. 3 generations of leptoquark transforming like the SM fermions). Discussing full models is beyond the scope of the present paper, but we illustrate some structures that can arise.

For triplet $\Delta_i$, the invariants can be quite different from those discussed in section \ref{sec:SU3}. If $\{ \Delta \}=0$ there is an invariant not involving the flavons $\epsilon^{ijk} \Delta_i Q_j L_k$.
This is a purely anti-symmetric structure for each $\lambda^i$:
\begin{equation}
\label{eq:asym_TD}
\lambda^1 = \lambda_0
\left( 
\begin{array}{ccc}
0 & 0 & 0 \\
0 & 0 & 1 \\
0 & -1 & 0
\end{array}
\right) \,, \quad
\lambda^2 = \lambda_0
\left(\begin{array}{ccc}
0 & 0 & -1 \\
0 & 0 & 0 \\
1 & 0 & 0
\end{array} \right) \,, \quad
\lambda^3 = \lambda_0 \left(\begin{array}{ccc}
0 & 1 & 0 \\
-1 & 0 & 0 \\
0 & 0 & 0
\end{array} \right) \,.
\end{equation}
For this charge assignment there are no holomorphic leptoquark bi-linears to study, but we can nevertheless conclude that this type of structure can not account for $R_K$ regardless of the mass eigenstates.

If $\{ \Delta \} \neq 0$, invariants can arise by contracting to 3 flavons, such as $(\phi_{23}^i \Delta_i) (\phi_{23}^j Q_j) (\phi_{23}^k L_k)$:
\begin{equation}
\label{eq:23_TD}
\lambda^1 = \lambda_0
\left( 
\begin{array}{ccc}
0 & 0 & 0 \\
0 & 0 & 0 \\
0 & 0 & 0
\end{array}
\right) \,, \quad
\lambda^2 = - \lambda^3= \lambda_0 b^3
\left(
\begin{array}{ccc}
0 & 0 & 0 \\
0 & 1 & -1 \\
0 & -1 & 1
\end{array} \right) \,.
\end{equation}
which occurs for $\{ \Delta \}=-3$, where $\Delta_1$ decouples and $\Delta_{2,3}$ couple with $\lambda^{2,3} \sim \lambda^{[-2]}$,  Eq.~(\ref{eq:D-2}). For this charge assignment, there is a holomoprhic bi-linear $(\phi_3^i \Delta_i) (\phi_{23}^j \Delta_j) + (\phi_{23}^i \Delta_i) (\phi_{3}^j \Delta_j)$. It could give rise to a mass contribution after the R-symmetry is broken:
\begin{equation}
\label{eq:M_TD}
M_\Delta = m_\Delta a b
\left( 
\begin{array}{ccc}
0 & 0 & 0 \\
0 & 0 & 1 \\
0 & 1 & - 2 
\end{array}\right) \, .
\end{equation}
Diagonalizing this mass matrix reveals $\Delta_1$ has vanishing mass prior to soft terms, so we expect it to remain as the lightest leptoquark - however it also has no coupling to SM fermions. The next lightest mass eigenstate with mass mass $(-1+\sqrt{2}) m_\Delta a b$, and the heaviest with mass $(1+\sqrt{2}) m_\Delta a b$, are respectively dominantly $\Delta_2$ (with some $\Delta_3$) and the orthogonal combination. As both $\Delta_{2,3}$ couple to SM fermions with the $\lambda^{[-2]}$ texture, this option reduces to a more involved version of the singlet leptoquark discussed before.

For anti-triplet $\Delta^i$, the invariants require only a single flavon contraction (contrasting to fermion mass structures that required 2, and the anti-triplet which as we have seen requires either 0 or 3).
An example would be for $\{ \Delta \} = -1$, where we have $(\Delta^i Q_i) (\phi_{23}^j L_j) + (\phi_{23}^i Q_i) (\Delta^j L_j)$:
\begin{eqnarray}
\lambda_1 = \lambda_0 b
\left( 
\begin{array}{ccc}
0 & 1 & -1 \\
1 & 0 & 0 \\
-1 & 0 & 0
\end{array}
\right)
\,, \quad
\lambda_2 = \lambda_0 b
\left( 
\begin{array}{ccc}
0 & 0 & 0 \\
0 & 2 & -1 \\
0 & -1 & 0
\end{array}
\right) \,, \quad
\lambda_3 = \lambda_0 b
\left( 
\begin{array}{ccc}
0 & 0 & 0 \\
0 & 0 & 1 \\
0 & 1 & -2
\end{array}
\right)
\, .
\end{eqnarray}
There is no contribution from holomorphic bi-linears, as $\epsilon_{ijk} \phi_{23}^i \Delta^j \Delta^k$ vanishes. We can still conclude that these structures would allow for $R_K \neq 1$ as long there are sufficiently light eigenstates containing $\Delta^{1,2}$.

\subsection{$A_4$}

If we have 3 generations of leptoquarks as an $A_4$ triplet $\Delta_i$ there can be invariant contractions
$\left[ L \Delta \right]= L_1 \Delta_1 + L_2 \Delta_3 + L_3 \Delta_2$, if under $Z_3$, $\{ \Delta \}=2$; within $A_4 \times Z_4$  the choice would be $\{ \Delta \}=3$. 
Each generation has its own $\lambda^i$ matrix. The LO-structures are:
\begin{equation}
\label{eq:A4_trip}
\lambda^1 = \lambda_0
\left( 
\begin{array}{ccc}
x_d & 0 & 0 \\
x_s & 0 & 0 \\
x_b & 0 & 0
\end{array}
\right) \,, \quad
\lambda^2 = \lambda_0
\left(\begin{array}{ccc}
0 & 0 & x_d \\
0 & 0 & x_s \\
0 & 0 & x_b
\end{array} \right) \,, \quad
\lambda^3 = \lambda_0 \left(\begin{array}{ccc}
0 & x_d & 0 \\
0 & x_s & 0 \\
0 & x_b & 0
\end{array} \right) \,.
\end{equation}
The holomorphic bi-linears in $\Delta$ allowed by $A_4\times Z_3$ are
$x_\xi \xi \left[ \Delta \Delta \right]$, $x_\xi' \xi' \left[ \Delta \Delta \right]''$ and $x_\nu \left[ \phi_\nu\Delta \Delta \right]$. This could lead to a mass structure 
\begin{align}
  M_{\Delta} &=  x_{\xi}
  \begin{pmatrix}
    1 & 0 & 0 \\
    0 & 0 & 1 \\
    0 & 1 & 0
  \end{pmatrix}
  + x_{\xi'}
  \begin{pmatrix}
    0 & 0 & 1 \\
    0 & 1 & 0 \\
    1 & 0 & 0
  \end{pmatrix}
  + \frac{1}{3} x_{\nu}
  \begin{pmatrix}
    2 & -1  & -1  \\
    -1  & 2 & -1  \\
    -1  & -1  & 2
  \end{pmatrix} \,,
  \label{eq:MD_A4T}
\end{align}
where we absorbed the magnitudes of the respective VEVs into the coefficients $x$.
This is similar to the Majorana neutrino structure in this framework (see \cite{Varzielas:2012ai}). With eigenvalue $x_\xi + x_\xi'$, $(1,1,1)$ is an eigenstate of this structure which, when compared to the respective $\lambda^i$, would preserve lepton universality.
In the limit $\langle \xi' \rangle=0$ we have effectively $x_\xi'=0$ and the other two eigenstates are also independent of the free parameters: they would be $(2,-1,-1)$ and $(0,1,-1)$, respectively with eigenstates $x_\nu + x_\xi$ and $x_\nu - x_\xi$.
These two leptoquark eigenstates could mediate LNU couplings.
As it is $\langle \xi' \rangle \neq 0$ that generates non-vanishing reactor angle in this $A_4 \times Z_3$ framework, in realistic regions of parameter space the $(2,-1,-1)$ and $(0,1,-1)$ directions are no longer eigenstates of the leptoquarks. $R_K \neq 1$ is still possible and an interesting situation arises where the perturbations away from the $(2,-1,-1)$ and $(0,1,-1)$ directions to the leptoquark mass eigenstates are directly related to $\theta_{13} \neq 0$ and the required  perturbations of the neutrino eigenstates, appearing in the leptonic mixing matrix (recall in this basis the charged leptons are diagonal).

Studying NLO corrections for the $A_4$ triplet $\Delta_i$, which requires correcting both the $\lambda^i$ and the assumed structures for $M_\Delta$, is beyond the scope of the present work.

\section{$d^c$ as $A_4$ triplet \label{app:tdc}}

We consider here a situation where $Q$ remains as singlets under the flavor symmetry but $d^c$ is, like $L$, an $A_4$ triplet. This situation arises naturally in $A_4$ unified models of lepton mixing \cite{deMedeirosVarzielas:2006fc, Altarelli:2008bg} with $SU(5)$.
Viable down quark masses can be obtained with LO contractions to the $\phi_l$ flavon, 
with NLO corrections from non-trivial singlet flavons and the up sector enabling viable CKM mixing.

If $d^c \sim 3$ one can have a RL leptoquark coupling with no flavon, $[d^c L] \lambda^{[]} \Delta$, which depending on $\Delta \sim 1, 1', 1''$ gives respectively $\lambda^{[]}$ structures:
\begin{align}
\lambda^{[]}=
  \lambda_{0}
  \begin{pmatrix}
    1 & 0 & 0  \\
    0 & 0 & 1  \\
    0 & 1 & 0
  \end{pmatrix} \,, \quad
\lambda^{[]}=\lambda_{0}
  \begin{pmatrix}
    0 & 0 & 1  \\
    0 & 1 & 0  \\
    1 & 0 & 0
  \end{pmatrix} \,, \quad
\lambda^{[]}=\lambda_{0}
  \begin{pmatrix}
    0 & 1 & 0  \\
    1 & 0 & 0  \\
    0 & 0 & 1
  \end{pmatrix} \,.
\end{align}
This structure would appear for $\{ \Delta \}=2$ within $A_4 \times Z_3$  and for  $\{ \Delta \}=3$  within $A_4 \times Z_4$.

Another possibility would be to couple to $\phi_l$,  $[d^c L]_{a,s} \phi_l \lambda^{[l]} \Delta$. For a $(u,0,0)$ VEV and $\Delta \sim 1, 1', 1''$, or alternatively for a $(0,u,0)$ VEV and $\Delta \sim 1'', 1, 1'$, these structures correspond respectively to $\lambda^{[l]}$:
\begin{align}
  \begin{pmatrix}
    2 a_{s} & 0 & 0  \\
    0 & 0 & a_{a}+a_{s}  \\
    0 & -a_{a}+a_{s} & 0
  \end{pmatrix} \,, \quad
  \begin{pmatrix}
    0 & a_{a}+a_{s} & 0  \\
    -a_{a}+a_{s} & 0 & 0  \\
    0 & 0 & 2 a_{s} 
  \end{pmatrix} \,, \quad
  \begin{pmatrix}
    0 & 0 & -a_{a}+a_{s} \\
    0 & 2 a_{s} & 0  \\
    a_{a}+a_{s} & 0 &  
  \end{pmatrix} \,.
\end{align}
One of these $\lambda^{[l]}$ can appear alone for the $A_4 \times Z_4$ framework and $\{ \Delta \}=1$, as $\{ \phi_l \}=2$. In the $A_4 \times Z_3$ framework $\{ \phi_l \}=0$ so the $\lambda^{[l]}$ structure appears for $\{ \Delta \}=2$, i.e. together with the respective $\lambda^{[]}$ structure for each choice $\Delta \sim 1, 1', 1''$ - the choice $1'$ having  $\lambda_{se,be} \neq 0$ and the choice $1''$ having $\lambda_{s\mu,b\mu} \neq 0$, these are two more candidates that can justify $R_K\neq1$.

The final possibility is to couple to $\phi_\nu$, with $(1,1,1)$ VEV, through $[d^c L]_{a,s} \phi_\nu \lambda^{[\nu]} \Delta$. $\Delta \sim 1, 1', 1''$ correspond to
\begin{align}
\lambda^{[\nu]}=
  a_{a}
  \begin{pmatrix}
    0 & +1  & -1  \\
    -1  & 0 & +1  \\
    1  & -1  & 0
 \end{pmatrix}
+  a_{s}
  \begin{pmatrix}
    2 & -1  & -1  \\
    -1  & 2 & -1  \\
    -1  & -1  & 2
  \end{pmatrix} \,.
\end{align}
In the $A_4 \times Z_4$ framework $\{ \phi_\nu \}=0$ so the $\lambda^{[\nu]}$ structure appears for $\{ \Delta \}=3$, i.e. together with the respective $\lambda^{[]}$ structure for each choice $\Delta \sim 1, 1', 1''$.
$\lambda^{[\nu]}$ can appear for the $A_4 \times Z_3$ framework and $\{ \Delta \}=0$, as $\{ \phi_\nu \}=2$, but this also allows $\Delta$ to couple to $\xi'$. Effectively this means that in both frameworks one must combine $\lambda^{[\nu]}$ with one of the three $\lambda^{[]}$.

\end{document}